\documentclass[runningheads]{llncs}
\usepackage[T1]{fontenc}
\usepackage{graphicx}
\usepackage{xspace}
\usepackage{lscape}
\usepackage[linesnumbered,vlined,ruled]{algorithm2e}
\usepackage{algpseudocode}
\SetKwInOut{KIN}{Input \ }
\SetKwInOut{KOUT}{Output \ }
\usepackage{amsmath,amssymb}
\usepackage{array,multirow,booktabs}
\usepackage{subfigure}
\usepackage{rotating}
\usepackage{supertabular}
\usepackage{wrapfig}
\usepackage{tcolorbox}
\tcbuselibrary{breakable}
\usepackage{xcolor}
\usepackage{soul}

\newtheorem{que}{\bf RQ}
\newtcolorbox{llmthought}[1][]{
    colback=gray!20,
    colframe=blue!60,
    boxsep=2pt,
    left=2pt,
    right=2pt,
    top=2pt,
    bottom=2pt,
    fontupper=\scriptsize,
    title=#1
}
\newtcolorbox{llmprompt}[1][]{
    colback=gray!0,
    colframe=black!100,
    boxsep=2pt,
    left=2pt,
    right=2pt,
    top=2pt,
    bottom=2pt,
    fontupper=\scriptsize,
    title=#1
}

\newcommand{\ie}{{\it i.e.}\xspace}

\newcommand{\eg}{{\it e.g.}\xspace}

\newcommand{\Solver}{\texttt{AutoICE}\xspace}
\newcommand{\ZeroShot}{\texttt{ZS}\xspace}
\newcommand{\FineTuneF}{\texttt{FT}-\texttt{fma}\xspace}
\newcommand{\FineTuneUF}{\texttt{FT}-\texttt{ultrachat}-\texttt{fma}\xspace}
\newcommand{\FineTuneTF}{\texttt{FT}-\texttt{tulu}-\texttt{fma}\xspace}
\newcommand{\LLMVer}{\texttt{LLMVer}\xspace}

\newcommand{\Ind}{I} 
\newcommand{\IndCode}{c_{\mathrm{ind}}} 
\newcommand{\IndScoreSyn}{s_{\mathrm{syn}}} 
\newcommand{\IndInfoSyn}{info_{\mathrm{syn}}} 
\newcommand{\IndScoreSem}{s_{\mathrm{sem}}} 
\newcommand{\IndInfoSem}{info_{\mathrm{sem}}} 
\newcommand{\CCode}{c_{\mathrm{C}}} 
\newcommand{\ACSLCode}{c_{\mathrm{ACSL}}} 
\newcommand{\NLRequirement}{r_{\mathrm{NL}}} 
\newcommand{\Population}{pop} 
\newcommand{\SizePop}{S_{\mathrm{P}}} 
\newcommand{\parent}{par} 
\newcommand{\offspring}{off} 
\newcommand{\MaxCall}{call_{\mathrm{max}}} 

\newcommand{\FMALPACA}{\textsf{FM}-\textsf{ALPACA}\xspace}
\newcommand{\FMBENCH}{\textsf{FM}-\textsf{BENCH}\xspace}
\newcommand{\FMALPACADF}{\textsf{FM}-\textsf{ALPACA}-\textsf{DF}\xspace}
\newcommand{\FMBENCHDF}{\textsf{FM}-\textsf{BENCH}-\textsf{DF}\xspace}

\newcommand{\fitness}{\mathsf{fitness}} 

\newcommand{\InitializePopulationLLM}{\textsc{initializePopulationLLM}\xspace} 
\newcommand{\CrossoverLLM}{\textsc{crossoverLLM}\xspace} 
\newcommand{\MutateLLM}{\textsc{mutateLLM}\xspace} 

\newcommand{\SizeIniPop}{\mathrm{S}_{\mathrm{IP}}} 
\newcommand{\NumElite}{\mathrm{\#elite}} 
\newcommand{\MutateR}{\mathrm{\rho}} 
\newcommand{\MaxGen}{\mathrm{\#gen}} 
\newcommand{\MaxIter}{\mathrm{\#iter}} 

\newcommand{\set}[1]{\left\{ #1 \right\}}

\newcommand{\tuple}[1]{\left( #1 \right)}

\newcommand{\code}[1]{\texttt{#1}}

\definecolor{rolecol}{HTML}{2B5F9E}      
\definecolor{taskcol}{HTML}{2E8B57}      
\definecolor{constraintcol}{HTML}{D35400} 
\definecolor{inputcol}{HTML}{6E6E6E}     
\definecolor{domaincol}{HTML}{6A1B9A}    

\newcommand{\RoleDef}[1]{\textcolor{rolecol}{#1}}
\newcommand{\TaskDesc}[1]{\textcolor{taskcol}{#1}}
\newcommand{\DomainDesc}[1]{\textcolor{domaincol}{#1}}
\newcommand{\TaskConstr}[1]{\textcolor{constraintcol}{#1}}
\newcommand{\InputText}[1]{\textcolor{inputcol}{#1}}
\begin{document}
\title{\Solver: Automatically Synthesizing Verifiable C Code via LLM-driven Evolution}
%
%
\author{Weilin Luo\orcidID{0000-0002-3733-9361} \and
Xueyi Liang\orcidID{0009-0003-8661-8396} \and
Haotian Deng\orcidID{0009-0007-7010-8278} \and
Yanan Liu\orcidID{0000-0001-7357-1793} \and
Hai Wan\orcidID{0000-0001-5357-9130}}
\authorrunning{W. Luo et al.}
%
\institute{Sun Yat-sen University, School of Computer Science and Engineering, Guangzhou, China\\
\email{\{luowlin5,wanhai\}@mail.sysu.edu.cn}, \email{\{liangxy233,denght7,liuyn56\}@mail2.sysu.edu.cn}}
\maketitle              
\begin{abstract}
Automatically synthesizing verifiable code from natural language requirements ensures software correctness and reliability while significantly lowering the barrier to adopting the techniques of formal methods. 
With the rise of large language models (LLMs), long-standing efforts at autoformalization have gained new momentum.
However, existing approaches suffer from severe syntactic and semantic errors due to the scarcity of domain-specific pre-training corpora and often fail to formalize implicit knowledge effectively. 
In this paper, we propose \Solver, an LLM-driven evolutionary search for synthesizing verifiable C code. 
It introduces the diverse individual initialization and the collaborative crossover to enable diverse iterative updates, thereby mitigating error propagation inherent in single-agent iterations. 
Besides, it employs the self-reflective mutation to facilitate the discovery of implicit knowledge.
Evaluation results demonstrate the effectiveness of \Solver: it successfully verifies $90.36$\% of code, outperforming the state-of-the-art (SOTA) approach. 
Besides, on a developer-friendly dataset variant, \Solver achieves a $88.33$\% verification success rate, significantly surpassing the $65$\% success rate of the SOTA approach.
\end{abstract}

\section{Introduction}\label{sec:introduction}

Autoformalization, the process of automatically translating informal requirements, \eg, natural language, into formal, machine-verifiable specifications, \eg, ANSI/ISO C specification language (ACSL)~\cite{Baudin2021acsl}, plays an increasingly pivotal role in formal methods.
Due to the widespread use of C code, one of the practical and impactful directions of autoformalization is the automated synthesis of verifiable C code, namely ACSL-annotated code.
Unlike standard code generation, synthesizing verifiable C code requires generating not only C code but also the corresponding formal specifications, \eg, pre-conditions, post-conditions, and loop invariants. 
By leveraging mature symbolic solvers, the correctness of the generated code can be mathematically proven.
However, manually writing formal specifications is a creative process that requires developers to invest a lot of extra effort, experience, and expertise, because formal specification languages are closer to mathematics than traditional programming languages~\cite{MurrayMBGBSLGK13,Leroy09,FariaA23,NobleSGS22}.
The automated generation of high-quality, verifiable C code is crucial and urgent.
It would significantly lower the barrier to entry for formal methods, which is of paramount importance for safety-critical domains such as finance, healthcare, and autonomous systems~\cite{Knight02,KleinAEMSKH14,Leroy09}.

However, achieving high-quality autoformalization is challenging.
    {\em Semantic gap}: natural language is ambiguous, polysemous, and expressive, whereas formal languages require absolute precision and singularity.
    {\em Data scarcity}: high-quality pairs of informal requirements and formal specifications are extremely rare, hindering the development of learning-based approaches~\cite{LiPP24,WuC0W0M24,MisuLM024,Cao0LMLH000QCT25}.

With the rapid advancement of large language models (LLMs), their capabilities in natural language understanding and code generation have garnered significant attention~\cite{BrownMRSKDNSSAA20,ChangWWWYZCYWWYZCYYX24,ZengMYZ24}. 
Trained on ultra-large-scale text and code, LLMs offer a promising new avenue for autoformalization~\cite{abs-2107-03374,LuGRHSBCDJTLZSZ21}.
LLMs demonstrate the ability to handle the semantic gap based on context.
For example, Wu et al.~\cite{WuJLRSJS22} successfully utilized LLMs to translate mathematical competition problems into Isabelle/HOL, a task previously considered intractable.
Cosler et al.~\cite{CoslerHMST23} proposed a tool to interactively translate temporal properties in natural language to temporal logics with LLMs.

Building on this potential, Cao et al.~\cite{Cao0LMLH000QCT25} have explored synthesizing verifiable C code via LLMs. 
Despite some progress, existing LLM-based approaches still face severe limitations.
\begin{itemize}
    \item Due to data scarcity and inherent hallucinations, LLMs frequently generate code with syntactic or semantic errors. 
    While feedback from verifiers can signal verification failures, this feedback is often sparse. 
    Furthermore, relying solely on LLMs to self-reflection code based on sparse feedback can lead to error accumulation rather than resolution.
    \item Due to variations in expertise among developers and the differences between formal languages and natural language, the requirements in natural language are often incomplete, making it difficult to formalize implicit knowledge.
    Natural language often omits implicit knowledge, \eg, common-sense background knowledge and intermediate reasoning steps, yet formal systems strictly require complete logical chains to verify correctness.
\end{itemize}

\begin{figure}[t]
    \centering
    \includegraphics[width=1\textwidth]{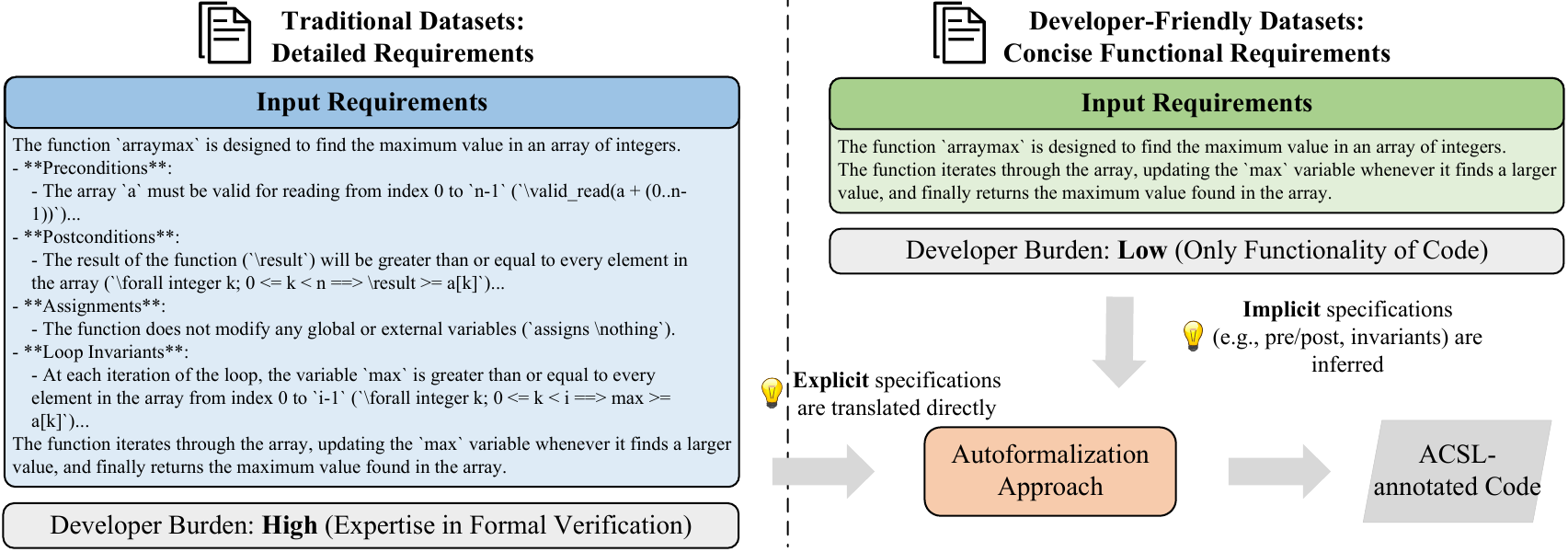}
    \caption{
        Traditional requirements vs. developer-friendly requirements.
    }\label{fig:setting}
\end{figure}

Besides, current works often predicate their success on the availability of overly rich requirement descriptions. 
For example, Cao et al.~\cite{Cao0LMLH000QCT25} assume that the input requirements characterize not only the functionality of code but also provide explicit details regarding pre-conditions, post-conditions, and loop invariants.
However, obtaining such detailed natural language requirements is challenging and imposes an additional cognitive burden on developers, requiring expertise they may not possess. 
A developer-friendly input should ideally consist solely of the functionality of code, leaving the complex formal intricacies to be inferred by the automation tool.
Fig.~\ref{fig:setting} illustrates the difference between traditional requirements and developer-friendly requirements.

To address these challenges, we propose \Solver, a novel framework that leverages LLM-driven evolution to enhance the synthesis of verifiable C code. 
We model the synthesis task as the evolution of a code population, where each individual represents a candidate code.
By integrating evolutionary search with logical verifiers, \Solver effectively strikes a balance between exploration and exploitation. 
The exploration capability empowers LLMs to uncover implicit knowledge, while the exploitation capability facilitates the progressive generation of high-quality code.
Specifically, \Solver employs a diverse initialization strategy to guide the synthesis of diverse code as a population via diverse thinking. 
Intuitively, a diverse population provides a rich reservoir of implementation strategies. 
This heterogeneity increases the likelihood of capturing implicit knowledge and generating syntactically and semantically correct code.
During evolutionary search, we introduce a cooperative crossover, which allows the insight of offspring synthesis to come from the comparison of parents, extracting valuable experience from parents to correct syntactic and semantic errors.
Furthermore, to handle incomplete requirements, we design a self-reflective mutation that encourages LLMs to introduce new insights and discover implicit knowledge.

We assess \Solver on $68$ instances across two datasets. 
To ensure a comprehensive evaluation, we conduct experiments across a diverse set of LLM backbones, encompassing both open-source and closed-source models with varying parameter scales. 
We compare \Solver against two state-of-the-art (SOTA) approaches: zero-shot prompting and fine-tuning approaches. 
Additionally, we introduce a strong LLM-verifier interactive baseline, where an LLM iteratively refines code based on feedback loops with verifiers.
The results demonstrate that \Solver achieves a consistent and significant improvement in the number of verifiable codes across all tested LLM backbones. 
In the best-performing scenario, \Solver successfully verifies $90.36$\% of the instances, whereas the leading SOTA approach achieves only $85.36$\%, representing a substantial improvement of $5$\%.
To further demonstrate practicality, we construct a developer-friendly variant of the datasets by stripping away auxiliary specifications, such as pre/post-conditions and loop invariants, retaining only functional descriptions. 
On the developer-friendly variants, the performance gap between \Solver and SOTA approaches widens (the average improvement increased from $15$\% to $26.66$\%). 
This result substantiates that our LLM-driven evolution is particularly effective at uncovering the implicit knowledge required for verification. 
Finally, ablation studies confirm the essential contributions of diverse individual initialization, collaborative crossover, the self-reflective mutation to the framework's success.


\section{Preliminaries}\label{sec:preliminary}

\subsection{ANSI/ISO C Specification Language}

The ANSI/ISO C specification language (ACSL)~\cite{Baudin2021acsl} is a behavioral specification language designed for formally annotating C code. 
An ACSL specification consists of several clause types that define the contract a function must fulfill. 
Common clause types include the \textit{requires}, \textit{assigns}, and \textit{ensures} clauses.
The \textit{requires} clause states the preconditions that must hold before the function is executed. 
These conditions often involve constraints on input parameters and the validity of memory references. 
The \textit{assigns} clause specifies which memory locations the function is permitted to modify during its execution. 
Besides, the \textit{ensures} clause defines the postconditions that must be satisfied after the function completes, typically relating the output state to the inputs.
The code that successfully verifies against its ACSL specification is guaranteed to adhere to the specified behavior. 
In addition to the three clauses, ACSL supports a variety of other clause types, which are detailed in its official documentation\footnote{The official documentation of ACSL is publicly available at https://frama-c.com/html/acsl.html.}.

\subsection{Frama-C and Its Weakest Precondition Plugin}

As a platform for the static analysis and formal verification of C code, Frama-C~\cite{CuoqKKPSY12,KirchnerKPSY15} has been used in safety-critical projects~\cite{BritoP10,Dordowsky15,UngAGLNP24}.
Frama-C can be used for syntax checking and basic validation of ACSL, \eg, wrong order of clauses. 
To perform formal verification, the weakest precondition (WP) plugin of Frama-C can check whether the code behaviors conform to ACSL specifications and prove the properties of the code.
Specifically, proof obligations are generated by the WP plugin of Frama-C, grounded in the principles of Hoare logic~\cite{Hoare69}.
Then the Why platform~\cite{Filliatre2003multi} translates these obligations into logical goals.
To prove the generated logical goals, Frama-C relies on automated theorem provers like Alt-Ergo~\cite{Conchon2006Ergo}, CVC4~\cite{BarrettCDHJKRT11}, and Z3~\cite{MouraB08} or on interactive proof assistants like Coq~\cite{CoqRefManual8.4}.
These provers attempt to verify these logical goals within specified limits. 
If a goal cannot be proven within the given limits, the plugin provides feedback to assist developers in refining either the code or its specification.
A C code is considered formally verified with respect to its ACSL specification only after all goals generated by the WP plugin have been successfully proven.

\subsection{Large Language Models}

With a large number of parameters, large language models (LLMs) are machine learning models trained on a vast corpus of data. 
Built on the Transformer architecture~\cite{VaswaniSPUJGKP17}, SOTA LLMs employ decoder-only models~\cite{VaswaniSPUJGKP17} to generate text. 
Specifically, LLMs continuously predict the next token and append it to the input sequence until a stop token is predicted. 
The generation process enables LLMs to produce coherent and contextually relevant text. 
Decoder-only models have demonstrated strong performance on code-related tasks~\cite{abs-2107-03374}, such as code generation, comprehension, summarization, and transformation~\cite{FirstRRB23,LiLZFCLC23,WuJLRSJS22}.

To effectively utilize LLMs for specialized tasks, \eg, autoformalization, prompt engineering~\cite{BachSYWRNSKBFAD22,abs-2209-01390,DeckersFKPS0P23,HouDWLC22,JiangOTMDTC22,LiuC22a} is essential. 
Prompt refers to the crafted inputs that guide LLMs toward desired outputs. 
While LLMs support conversational interaction, applying LLMs to specialized tasks needs carefully structured prompts that align with both the task objectives and the used LLMs~\cite{Beurer-Kellner023,ReynoldsM21}. 
For example, program synthesis requires prompts to provide unambiguous task descriptions~\cite{BachSYWRNSKBFAD22,abs-2209-01390} and include examples that conform to the target programming language. 

\section{Related Works}\label{sec:related-work}

This section reviews existing works in the key areas: specification synthesis and the integration of LLMs with evolutionary search.

\subsection{Specification Synthesis}

Formal verification provides a rigorous approach for assuring that a system possesses critical properties. 
A key challenge is specification synthesis, \ie, generating a formal language for verification.
We summarize existing works by describing the transformation of different forms of input into formal languages.

\subsubsection{From Informal Language to Verifiable Formal Language.}

With the significant progress of LLMs, lots of works recently has focused on the transformation from informal requirements, such as natural language requirements, to verifiable formal languages.
Misu et al.~\cite{MisuLM024} explored the capability and potentiality of LLMs at synthesizing Dafny.
Lu et al.~\cite{LuSSHZJLT25} envisioned a framework that combines software engineering and formal approaches to achieve ACSL-annotated code from natural language requirements.
However, they do not provide the code, and the technical details are too vague to reproduce.
Cao et al.~\cite{Cao0LMLH000QCT25} further contributed to a comprehensive assessment of the LLMs' capability through considering five formal specifications, including Coq, Dafny, Lean4, ACSL, and TLA+.
However, it is confined to basic zero-shot prompting and fine-tuning, overlooking further avenues for enhancement. 
Additionally, their datasets contains requirement descriptions that are unrealistically detailed and thus hard to acquire in real-world scenarios.

We explore a high-performance specification synthesis approach for ACSL and expanded the datasets, hoping to promote the development of developer-friendly approaches.

\subsubsection{From Formal Language to Verifiable Formal Language.}

Generating verifiable formal languages from formal languages, \eg, JAVA and C code, has been extensively studied.
The verifiable formal languages of interest include, but are not limited to, 
class specifications~\cite{MolinadA23}, 
typestate specifications~\cite{BeckmanN11}, 
preconditions~\cite{RamanathanGJ07,CousotCFL13,PadhiSM16}, 
postconditions~\cite{PopeeaC06,SuAD18,SingletonLRC18}, 
loop invariants~\cite{DilligDLM13,LinZCSXLS21,Yu0023,RyanWYGJ20,SiDRNS18,YaoRWJG20,KarbyshevBIRS17,LinSXLSH17,0001BN24,PeiBSSY23,CaoWXYWCM25,YuLWW24,WuC0W0M24,PirzadaRBC24}, 
termination specification~\cite{LeQC15}, 
API specifications~\cite{ShohamYFP08}, 
Isabelle~\cite{LinCHWLLSL24,FirstRRB23}, and 
formal proof~\cite{YangSGCSYGPA23}. 

We, in particular, focuses on the synthesis of ACSL. 
Wen et al.~\cite{WenCSXQHLCT24} synthesized full ACSL specifications based on static analysis and program verification, including not only loop invariants but also preconditions, postconditions, and assigned variables.
Existing LLM-based approaches often struggle with programs containing complex loop structures, which can lead to the generation of irrelevant specifications. To address this issue, Chen et al.\cite{abs-2509-09917} proposed an approach tailored for such programs, utilizing program slicing and logical deletion.
Although the inputs to the works~\cite{WenCSXQHLCT24,abs-2509-09917} differ from the natural language requirements we are concerned with, their idea of combining LLMs with other techniques inspired us to combine LLMs with evolutionary search.

\subsubsection{From Multimodal Inputs to Verifiable Formal Language.}

In addition, a line of works focuses on synthesizing specifications from multimodal inputs, which combine both formal and informal languages. 
For instance, Zhai et al.~\cite{ZhaiSPZLFM0020} synthesized specifications from code and its natural language comments. 
Furthermore, leveraging LLMs, Mukherjee and Delaware~\cite{abs-2410-14835} generated Rocq/Coq from a combination of formal and informal specifications and test cases. 

\subsection{LLM-driven Evolutionary Search}\label{sec:related-work}

A recent emerging paradigm is LLM-based evolutionary search, which synergizes LLMs with evolutionary search to automatically generate heuristics. 
This powerful combination has found widespread application in areas including code generation~\cite{NejjarZSW25,HembergMO24} and text generation~\cite{Guo0GLS0L0Y24,abs-2406-07496}.
FunSearch~\cite{RomeraParedesBNBKDREWFKF24} first employs island-based evolution with LLMs to solve mathematical problems. 
Similarly, Lu et al.\cite{0044TY0LWL024} proposed EoH to combine genetic algorithms with chain-of-thought to tackle combinatorial optimization challenges. 
And ReEvo~\cite{Ye0CBHKPS24} introduces a reflective mechanism using dual LLMs to refine heuristics. 
More recently, Dat et al.\cite{DatDB25} proposed an adaptive LLM-based framework to maintain a balance between diversity and convergence with a harmony search algorithm. 
These approaches highlight the potential of integrating LLMs with evolutionary search.

\section{LLM-driven Evolution}\label{sec:method}

In this section, we present \Solver, a novel framework that leverages LLM-driven evolution to enhance the synthesis of verifiable C code.

\subsection{Overview}

\Solver takes natural language requirements as input and outputs an ACSL-annotated code. 
An vanilla approach can be repeated LLM generations with feedback coming from verifiers to search the code space, which suffers from error accumulation and lacks the ability to discover implicit knowledge.
Therefore, we propose three mechanisms to fully unleash the power of LLMs, namely diverse individual initialization, collaborative crossover, and self-reflective mutation.

\begin{algorithm}[h]
    \caption{\Solver}\label{alg:solver}
    \KIN{A natural language requirement $\NLRequirement$.}
    \KOUT{ACSL-annotated code or `$Not\ Synthesized$'.}
    
    $\Population \gets$ \Call{\InitializePopulationLLM}{$\NLRequirement$, $\SizeIniPop$}\label{alg:solver-InitializePopulationLLM}\\
    \If{There exists an individual $\hat{\Ind}  = \tuple{\hat{\IndCode}, \hat{\IndScoreSyn}, \hat{\IndInfoSyn}, \hat{\IndScoreSem}, \hat{\IndInfoSem} }$ in $\Population$ with $\fitness(\hat{\Ind}) = 2$}{
        \Return{$\hat{\IndCode}$}\\
    }
    \For{The generation from $1$ to $\MaxGen$}{\label{alg:solver-iter-begin}
        $\Population' \gets$ Select the top $\NumElite$ individuals from $\Population$\label{alg:solver-elite}\\
        \While{$|\Population'| < \SizePop$}{
            $\parent_1$, $\parent_2$ $\gets$ Randomly select two individuals from $\Population$ with the highest fitness value\\
            $\offspring_1$, $\offspring_2$ $\gets$ \Call{\CrossoverLLM}{$\parent_1$, $\parent_2$, $\NLRequirement$}\label{alg:solver-CrossoverLLM}\\
            \If{Randomly generate a number in $[0,1]$ $<$ $\MutateR$}{
                $\offspring_1$ $\gets$ \Call{\MutateLLM}{$\offspring_1$, $\NLRequirement$}\label{alg:solver-MutateLLM-1}\\
            }
            \If{Randomly generate a number in $[0,1]$ $<$ $\MutateR$}{
                $\offspring_2$ $\gets$ \Call{\MutateLLM}{$\offspring_2$, $\NLRequirement$}\label{alg:solver-MutateLLM-2}\\
            }
            $\Population'$ $\gets$ $\Population' \cup \set{\offspring_1, \offspring_2}$\\
        }
        $\Population \gets \Population'$\\
        \If{There exists an individual $\hat{\Ind}  = \tuple{\hat{\IndCode}, \hat{\IndScoreSyn}, \hat{\IndInfoSyn}, \hat{\IndScoreSem}, \hat{\IndInfoSem} }$ in $\Population$ with $\fitness(\hat{\Ind}) = 2$}{
            \Return{$\hat{\IndCode}$}\\
        }
    }\label{alg:solver-iter-end}
    \Return{`$Not\ Synthesized$'}\\
\end{algorithm}

Algorithm~\ref{alg:solver} outlines the overall workflow of \Solver.
Within the evolutionary search, LLMs play three roles: initializer, generating the initial population; crossover operator, producing offspring from parent pairs; and mutator, introducing mutations to generate variant offspring.
First, we initialize a population of size $\SizeIniPop$ based on diverse individual initialization (Section~\ref{sec:mithod-individual-initialization}), where $\SizeIniPop$ is a hyperparameter (line~\ref{alg:solver-InitializePopulationLLM}).
Then, we iteratively evolve the population through generations (lines~\ref{alg:solver-iter-begin}-\ref{alg:solver-iter-end}).
At each generation, the evolutionary search begins with the population generated by the previous generation, followed by three iterative steps: selection, crossover (Section~\ref{sec:method-crossover}), and randomly mutation with mutation rate $\MutateR$ (Section~\ref{sec:method-mutation}), where $\MutateR$ is a hyperparameter.
For selection, we randomly retain $\NumElite$ elite individuals from this generation, \ie, those with the highest fitness (Equ.~\eqref{equ:fitness}), to prevent the optimal solution from being lost (line~\ref{alg:solver-elite}).
Evolutionary search iterates continuously until it reaches the maximum generations $\MaxGen$ or finds code that can pass the WP plugin of Frama-C.
$\NumElite$ and $\MaxGen$ are hyperparameters.
Notably, we maintain a fixed population size $\SizePop$ based on $\SizeIniPop$ and $\NumElite$, as computed in Equ.~\eqref{equ:sizepop}.

\begin{equation}\label{equ:sizepop}
    \SizePop = \NumElite + \lceil(\SizeIniPop - \NumElite) / 2 \rceil \cdot 2
\end{equation}

\subsubsection{Individual encoding.}

We define an individual as ACSL-annotated code along with its verification-related information.
Formally, an individual is a $5$-tuple $\Ind = \tuple{\IndCode, \IndScoreSyn, \IndInfoSyn, \IndScoreSem, \IndInfoSem}$, where $\IndCode$ is an ACSL-annotated code, $\IndScoreSyn$ and $\IndScoreSem$ are whether $\IndCode$ passes Frama-C and the WP plugin of Frama-C, respectively, and $\IndInfoSyn$ and $\IndInfoSem$ are the hints output by Frama-C and the WP plugin of Frama-C, respectively.
If $\IndCode$ passes Frama-C, then $\IndScoreSyn = 1$; otherwise, $\IndScoreSyn = 0$; 
if $\IndCode$ passes the WP plugin of Frama-C, then $\IndScoreSem = 1$; otherwise, $\IndScoreSem = 0$.
Once $\IndCode$ changes, we will call the verifiers to update $\IndScoreSyn$, $\IndInfoSyn$, $\IndScoreSem$, and $\IndInfoSem$ synchronously.

\subsubsection{Fitness.}

The fitness of an individual is defined as:
\begin{equation}\label{equ:fitness}
    \fitness(\tuple{\IndCode, \IndScoreSyn, \IndInfoSyn, \IndScoreSem, \IndInfoSem}) = \IndScoreSyn + \IndScoreSem.
\end{equation}
The fitness value ranges from $0$ to $2$.
An individual with a fitness value of $1$ indicates that its code only passes Frama-C.
An individual with a fitness value of $2$ indicates that its code passes both Frama-C and the WP plugin of Frama-C, which is the optimal solution we aim to find. 
Once such an individual is found, the evolutionary search terminates immediately.

\subsection{Diverse Individual Initialization}\label{sec:mithod-individual-initialization}

Our insight stems from the `mental set' theory in psychology~\cite{Jersild27}, which has recently shown promise in enhancing LLM reasoning and debate~\cite{Liu0L0T25}. 
A mental set constitutes a cognitive bias that impedes divergent thinking. 
It traps individuals in habitual thinking patterns, making it difficult to complete tasks outside of established thinking patterns.
However, evidence suggests that breaking the mental set often reveals solutions that were previously obscured. 
Recently, Liu et al.~\cite{Liu0L0T25} have discovered that this theory also applies to LLMs.

Motivated by this, we employ diverse prompting reasoning strategies, \eg, chain-of-thought prompting (CoT)~\cite{Wei0SBIXCLZ22}, during initialization to facilitate the synthesis of differentiated code.
These strategies fully leverage the inherent capabilities of off-the-shelf LLMs without the need for training or fine-tuning, and can effectively represent distinct thinking patterns~\cite{Liu0L0T25}.
We prioritize selecting strategies with significant disparities to mitigate potential mental set issues arising from similar strategies.
Intuitively, divergent code enhances population diversity, yielding a broader array of C code and ACSL specifications, which increases the potential of generating syntactically and semantically correct code snippets or discovering novel implicit knowledge.

The initialization of an individual adopts a two-phase strategy designed to mirror the logical workflow of developers. 
In the first phase, we synthesize C code directly from the natural language requirements. 
In the second phase, we synthesize the corresponding ACSL specifications, utilizing both the original requirements and the code generated in the previous step as context.
The sequence of these phases is determined by the difficulty of the tasks. 
Due to the differences in the quantity and quality of pre-training data of LLMs, it is more challenging for LLMs to synthesize ACSL specifications than to synthesize C code~\cite{Cao0LMLH000QCT25}.
Therefore, providing C code as an intermediate reference serves to lower the barrier for the subsequent formalization of ACSL specifications.

\begin{figure}[t]
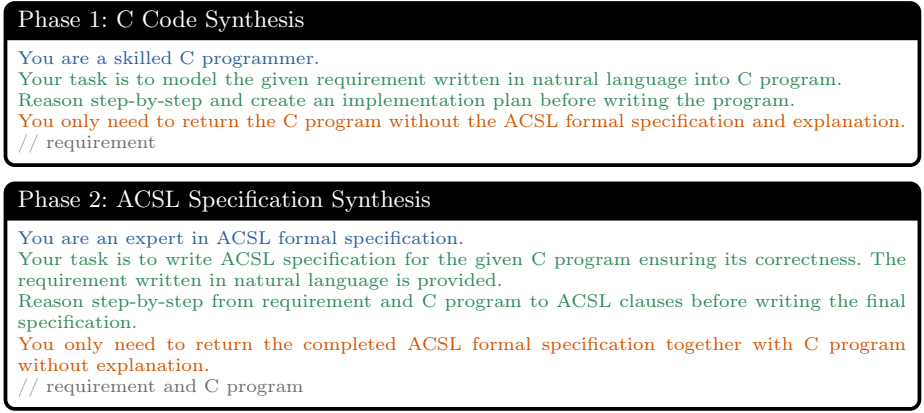

    \centering
    \begin{llmprompt}[Phase 1: C Code Synthesis]
        \RoleDef{You are a skilled C programmer.}

        \TaskDesc{Your task is to model the given requirement written in natural language into C program.}

        \TaskDesc{Reason step-by-step and create an implementation plan before writing the program.}

        \TaskConstr{You only need to return the C program without the ACSL formal specification and explanation.}

        \InputText{// requirement}
    \end{llmprompt}
    \begin{llmprompt}[Phase 2: ACSL Specification Synthesis]
        \RoleDef{You are an expert in ACSL formal specification.} 

        \TaskDesc{Your task is to write ACSL specification for the given C program ensuring its correctness. The requirement written in natural language is provided.}

        \TaskDesc{Reason step-by-step from requirement and C program to ACSL clauses before writing the final specification.}

        \TaskConstr{You only need to return the completed ACSL formal specification together with C program without explanation.}

        \InputText{// requirement and C program}
    \end{llmprompt}
    \caption{An example of the prompts for the two-phase, where 
    colors indicate role definition (\RoleDef{blue}), task description (\TaskDesc{green}), task constraint (\TaskConstr{orange}), and inputs (\InputText{gray}).
    We omit the inputs and mark them as `//'.}\label{fig:prompt-individual-initialization}
\end{figure}

Technically, each phase is driven by a prompt consisting of four components: role definition, task description, task constraint, and inputs. 
Fig.~\ref{fig:prompt-individual-initialization} shows an example of the prompts for the two-phase.
This prompt explicitly specifies the role of LLMs, defines the task objectives, and constrains the output format to guide the response of LLMs.

\subsection{Collaborative Crossover}\label{sec:method-crossover}

The efficacy of evolutionary search hinges on meticulously designed crossover operators that enable offspring to inherit superior genetic traits from their progenitors. 
To this end, we propose a collaborative crossover, designed to synthesize the strengths of parent individuals, \ie, correct syntax and semantics.
Specifically, we first randomly select parent pairs from the pool of individuals that have the highest fitness value.
The selection criterion aims to minimize the propagation of syntactically erroneous code fragments to the next generation. 
Subsequently, treating one parent as the baseline, we instruct LLMs to refine this baseline by incorporating insights and logic from the other parent.

Technically, the prompt of the collaborative crossover is structured into five components: role definition, domain description, task description, output constraint, and inputs.
Fig.~\ref{fig:prompt-crossover} illustrates an example of the crossover prompt.
This prompt instructs LLMs to analyze two candidate solutions and perform crossover-based code refinement by combining their most advantageous characteristics.

\begin{figure}[t]
    \centering
    \begin{llmprompt}[Collaborative Crossover]
        \RoleDef{You are an expert in C code optimization and formal verification with ACSL.} 

        \DomainDesc{The requirement written in natural language is provided. Below are two individuals implemented based on this requirement, parent1 and parent2, each with the following attributes:}

        \DomainDesc{- full\_code: C code with ACSL-annotations}

        \DomainDesc{- base\_info: Frama-C verification report without WP plugin}

        \DomainDesc{- base\_pass: Whether the base Frama-C verification passed}

        \DomainDesc{- wp\_info: Frama-C verification report with WP plugin}

        \DomainDesc{- wp\_pass: Whether the WP-based verification passed}

        \TaskDesc{Your task is to analyze parent1 and parent2 comparatively, focusing on:}
        
        \TaskDesc{- Consistency between the full\_code and requirement}
        
        \TaskDesc{- ACSL annotation completeness and correctness}
        
        \TaskDesc{- Reasons for the success or failure of base and WP verification}
        
        \TaskDesc{- Overall code robustness and specification adherence}
        
        \TaskDesc{Then, propose a refined full\_code of parent1 that:}
        
        \TaskDesc{- Incorporates strengths from parent2 (e.g., better annotations, verified constructs)}
        
        \TaskDesc{- Addresses weaknesses in parent1 identified via Frama-C reports}
        
        \TaskDesc{- Enhances fitness by improving verification outcomes}
        
        \TaskDesc{- Maintains functional correctness and clarity}

        \TaskConstr{You only need to return the refined full\_code of parent1 without explanation.}

        \InputText{// requirement, parent1's attributes, and parent2's attributes}
    \end{llmprompt}
    \caption{An example of the crossover prompt and a schematic diagram of the collaborative crossover, where colors indicate domain description (\DomainDesc{purple}).}\label{fig:prompt-crossover}
\end{figure} 

\subsection{Self-reflective Mutation}\label{sec:method-mutation}

In evolutionary search, the mutation plays a pivotal role in encouraging offspring to transcend parental constraints, thereby facilitating the escape from local optima. 
Accordingly, we design a self-reflective mutation with the objective of enabling offspring to rectify syntactic and semantic errors or discover implicit knowledge through a process of self-reflection.
Unlike the crossover, which relies on parents, the self-reflective mutation drives LLMs to critically analyze the code and generate variations based on verifier feedback. 

Technically, the prompt of the self-reflective mutation is similarly composed of a role definition, domain description, task description, output constraint, and inputs. 
Fig.~\ref{fig:prompt-mutation} also provides a representative example of the self-reflective mutation prompt.
By analyzing verification reports and requirements, it must autonomously diagnose and resolve inconsistencies, simulating an analytical process of an expert to ensure correctness.

\begin{figure}[t]
    \centering
    \begin{llmprompt}[Self-reflective Mutation]
        \RoleDef{You are an expert in C code optimization and formal verification with ACSL.}

        \DomainDesc{The requirement written in natural language is provided. Below is an individual implemented based on this requirement, with the following attributes:}

            \DomainDesc{- full\_code: C code with ACSL annotations}

            \DomainDesc{- base\_info: Frama-C verification report without WP plugin}

            \DomainDesc{- base\_pass: Whether the base Frama-C verification passed}

            \DomainDesc{- wp\_info: Frama-C verification report with WP plugin}

            \DomainDesc{- wp\_pass: Whether the WP-based verification passed}
            
        \TaskDesc{Analyze the provided full\_code with its verification reports. Your goal is to generate an improved mutated version, focusing on:}

            \TaskDesc{- Consistency between the full\_code and requirement}

            \TaskDesc{- ACSL Annotations: Check completeness, correctness, and consistency with code behavior}

            \TaskDesc{- Verification Gaps: Identify why base\_pass or wp\_pass failed (if applicable)}
            
            \TaskDesc{- Code Logic: Review for potential bugs, inefficiencies, or specification mismatches}
            
        \TaskDesc{Then, propose a refined full\_code of the individual that:}

            \TaskDesc{- Ensure all Frama-C verification warnings are addressed}

            \TaskDesc{- Maintain functional correctness and code clarity}
            
        \TaskConstr{You only need to return the refined full\_code of the individual without explanation.}

        \InputText{// requirement and individual's attributes}
    \end{llmprompt}
    \caption{An example of the mutation prompt and a schematic diagram of the self-reflective mutation.}\label{fig:prompt-mutation}
\end{figure}

\section{Experimental Evaluation}\label{sec:experiment}

In this section, we conduct experiments to answer four research questions\footnote{Our code and datasets are publicly available at: https://github.com/lbearxy/AutoICE.}.

\begin{que}\label{que:experiment-1}
    What is the performance of \Solver?
\end{que}

\begin{que}\label{que:experiment-2}
    How effective is \Solver on developer-friendly datasets?
\end{que}

\begin{que}\label{que:experiment-3}
    Is every step of \Solver designed reasonably?
\end{que}

\begin{que}\label{que:experiment-4}
    How strong is the robustness to the hyperparameters of \Solver?
\end{que}

\subsection{Datasets}

\subsubsection{\FMALPACA and \FMBENCH.}

We adopt two datasets \FMALPACA and \FMBENCH constructed by Cao et al.~\cite{Cao0LMLH000QCT25}\footnote{The datasets are publicly available at https://huggingface.co/fm-universe/datasets.}. 
These datasets encompass five formal specification languages, \ie, Coq~\cite{Huet87}, Lean4~\cite{Moura021}, Dafny~\cite{Leino10}, ACSL~\cite{Baudin2021acsl,CuoqKKPSY12}, and TLA+~\cite{YuML99,Lamport2002}, across six formal-verification-related tasks, generated through distillation from GPT-4o~\cite{abs-2410-21276}.
We specifically focuses on ACSL and the task of formalizing natural language requirements into ACSL-annotated code. 
\FMALPACA contains $56$ instances, while \FMBENCH contains $14$. 
We exclude two instances from \FMBENCH as they involve lambda functions, which is not yet implemented in Frama-C.

\subsubsection{Developer-friendly Variants.}

The natural language requirements primarily include descriptions of the target function, preconditions, postconditions, loop annotations, and assertions. 
To address the lack of ACSL expertise among most developers, we extend the datasets to include only the description of the target function, called \FMALPACADF and \FMBENCHDF.

\subsection{Baselines}

We establish a comprehensive set of SOTA approaches that encompass three distinct paradigms. 
Following the prior work of Cao et al.~\cite{Cao0LMLH000QCT25}, we include the zero-shot and fine-tuning baselines to serve as fundamental approaches of comparison. 
Furthermore, to provide a more competitive comparison, we extend the baselines to include an LLM-verifier interaction approach.
The interaction paradigm demonstrates SOTA performance on other problems~\cite{LinCHWLLSL24,LuoFQWLY24,WuC0W0M24,SevenhuijsenEN25}.
We do not compare with ReDeFo~\cite{LuSSHZJLT25} because they only provide an envision and not the available code.

\subsubsection{Zero-shot Approach.}

The zero-shot approach, denoted by \ZeroShot, provides only a task instruction and inputs, instruct LLMs to generate an ACSL-annotated code immediately without any examples or training. 
We utilize the same prompt template as Cao et al.~\cite{Cao0LMLH000QCT25} to ensure a fair comparison.

\subsubsection{Fine-tuning Approach.}

We directly employ the fine-tuning LLMs provided by Cao et al.~\cite{Cao0LMLH000QCT25} with zero-shot prompting for baseline evaluation\footnote{The fine-tuning LLMs are publicly available at https://huggingface.co/fm-universe/models.}. 
It includes the instruction-tuned versions of llama3.1-instruct (8B)~\cite{Meta2024Llama3}, qwen2.5-instruct (7B)~\cite{abs-2412-15115}, qwen2.5-coder-instruct (7B)~\cite{abs-2409-12186}, and deepseek-coder-instruct (7B)~\cite{abs-2501-12948}, all of which were fine-tuning on \FMALPACA, denoted by \FineTuneF. 
Besides, to ensure a comprehensive comparison, llama3.1-instruct (8B)~\cite{Meta2024Llama3} was fine-tuned on two widely-used datasets: UltraChat~\cite{DingCXQHL0Z23} and Tulu-V3~\cite{abs-2411-15124} before fine-tuning on \FMALPACA, denoted by \FineTuneUF and \FineTuneTF, respectively.

\subsubsection{LLM-verifier Interaction Approach.}

\begin{figure}[t]
    \centering
    \begin{llmprompt}[Refinement]
        \RoleDef{You are an expert in C code optimization and formal verification with ACSL.}

        \DomainDesc{The requirement written in natural language is provided. The ACSL-annotated code you generated failed verification with error information reported by Frama-C.}

        \TaskDesc{Your task is to analyze the error messages and provide precise corrections to make the ACSL-annotated code verifiable.}
               
        \TaskConstr{You only need to return the completed ACSL formal specification with the code without explanation.}

        \InputText{// requirement, ACSL-annotated code and error messages}
    \end{llmprompt}
    \caption{An example of the refinement prompt.}\label{fig:prompt-refinement}
\end{figure}

Algorithm~\ref{alg:llm-verifier} shows the iterative refinement process of LLM-verifier interaction approach, denoted by \LLMVer. 
This approach establishes a self-correcting, verifier-driven closed-loop system based on a `generate–verify–correct' workflow.
For fairness, the initialization of Algorithm~\ref{alg:llm-verifier} (lines~\ref{alg:llm-verifier-step1}-\ref{alg:llm-verifier-step2}) uses the same prompts as the individual initialization of \Solver.
Fig.~\ref{fig:prompt-refinement} demonstrates the refinement prompt for LLMs (line~\ref{alg:llm-verifier-step3} and~\ref{alg:llm-verifier-step4}), which consists of five components: role definition, domain description, task specification, output constraint, and inputs.

\begin{algorithm}[h]
    \caption{\LLMVer}\label{alg:llm-verifier}
    \KIN{A natural language requirement $\NLRequirement$.}
    \KOUT{ACSL-annotated code or `$Not\ Synthesized$'.}

    $\CCode \gets$ Synthesize C code from $\NLRequirement$ via LLMs\label{alg:llm-verifier-step1}\\
    $\ACSLCode \gets$ Synthesize ACSL-annotated code from $\CCode$ and $\NLRequirement$ via LLMs\label{alg:llm-verifier-step2}\\
    $n_{\mathrm{iter}} \gets \MaxIter$\\
    
    \For{$n_{\mathrm{iter}} > 0$}{
        $err_{FC} \gets$ Verify $\ACSLCode$ using Frama-C\\
        
        \If{verification succeeds}{
            {\bf break}\\
        }\Else{
            $\ACSLCode \gets$ Refine $\ACSLCode$ based on $\NLRequirement$ and $err_{FC}$ via LLMs\label{alg:llm-verifier-step3}\\
            $n_{\mathrm{iter}} \gets n_{\mathrm{iter}} - 1$
        }
    }
    \For{$n_{\mathrm{iter}} > 0$}{
        $err_{wp} \gets$ Verify $\ACSLCode$ using the WP plugin of Frama-C\\
        
        \If{verification succeeds}{
            \Return{$\ACSLCode$}\\
        }\Else{
            $\ACSLCode \gets$ Refine $\ACSLCode$ based on $\NLRequirement$ and $err_{wp}$ via LLMs\label{alg:llm-verifier-step4}\\
            $n_{\mathrm{iter}} \gets n_{\mathrm{iter}} - 1$
        }
    }
    \Return{`$Not\ Synthesized$'}\\
\end{algorithm}

\subsection{Experimental Setups}

\subsubsection{Evaluation Metrics.}
Following prior works~\cite{Cao0LMLH000QCT25},
we adopt Pass@1, average Pass@1, and Pass@5 accuracy over five independent trials as the primary metrics.
\begin{itemize}
    \item Pass@1 measures the rate of ACSL-annotated code that pass formal verification on their first generation attempt. 
    It reflects the baseline accuracy of approaches in a one‑attempt setting.
    \item Average Pass@1 is computed as the mean of Pass@1 accuracy across five independent generation runs. 
    This metric smooths out variability caused by randomness in the generation process and offers a more stable estimate.
    \item Pass@5 measures the probability that at least one of five independently generated ACSL-annotated code passes formal verification. 
    It evaluates the ability to produce a valid specification within a bounded number of attempts.
\end{itemize}

\subsubsection{Configurations.}

Following prior works~\cite{Cao0LMLH000QCT25}, we conduct tests on $8$ LLM backbones with different parameter scales, specifically llama3.1-instruct (8B/70B)~\cite{Meta2024Llama3}, qwen2.5-instruct (7B/72B)~\cite{abs-2412-15115}, qwen2.5-coder-instruct (7B/32B)~\cite{abs-2409-12186}, deepseek-coder-instruct (7B)~\cite{abs-2501-12948}.
Besides, we conduct tests on two representative open-source LLMs, qwen3-AWQ (32B)~\cite{abs-2505-09388} and deepseek-R1 (671B)~\cite{abs-2501-12948}, with significantly different parameters, and one representative closed-source LLM, gemini-2.5-pro~\cite{abs-2507-06261}.
Given that \FMALPACA and \FMBENCH are generated by GPT-4o~\cite{abs-2410-21276}, we exclude LLMs from the GPT series to prevent potential evaluation bias.
For all LLMs, we set the temperature to $1$ and use the default values for other hyperparameters.
We select the following two widely used prompting reasoning strategies: CoT~\cite{Wei0SBIXCLZ22} and step-back prompting (SBP)~\cite{ZhengMCCCLZ24}.
For verifying the ACSL-annotated code, we employ Frama-C 31.0 (Gallium).
All experiments are run on a GPU server equipped with Intel(R) Xeon(R) Gold 6248R CPU @3.0GHz, 810 GB RAM and NVIDIA A100 GPU acceleration. 
The main hyperparameters of \Solver are set as follows: $\SizeIniPop = 5$, $\NumElite = 2$, $\MutateR = 0.5$, and $\MaxGen = 5$.
To ensure a fair comparison between \Solver and \LLMVer, we impose the same upper bound on the number of LLM calls, denoted as $\MaxCall$. 
The expected $\MaxCall$ of \Solver is computed as: $\mathbb{E}[\MaxCall] = 2 \cdot \SizeIniPop + \MaxGen \cdot (\SizePop - \NumElite) \cdot (1 + \MutateR \cdot 1)$, where $\SizePop$ is defined in Equ.~\eqref{equ:sizepop}. 
Specifically, $2 \cdot \SizeIniPop$ accounts for the two LLM calls per initial individual. $\SizePop - \NumElite$ is the number of new offspring created in each generation. 
Each offspring requires one call for crossover plus an additional call for mutation with probability $\MutateR$.
Therefore, we set the hyperparameter $\MaxIter$ of \LLMVer as \(\MaxIter = \mathbb{E}[\MaxCall] - 2\), since the initialization calls LLM for two times.
This constraint ensures that both approaches are evaluated under comparable computational limits.

\subsection{Result Analysis}

\begin{table}[htbp]
    \centering
    \caption{Average Pass@1 (\%) across different approaches, where 
    `Size' means the parameter size of LLMs, 
    `/' means the parameter size is not disclosed, 
    `FC' means the verifier is Frama-C, 
    `WP' means the verifier is the WP plugin of Frama-C, and 
    {\bf boldface} numbers refer to the better results for each LLM.
    The larger, the better.}\label{tab:result-ave-pass1}%
    \scalebox{0.9}{
        \begin{tabular}{c|c|l|rr|rr|rr|rr}
        \toprule
        \multirow{2}[2]{*}{LLMs} & \multirow{2}[2]{*}{Sizes} & \multicolumn{1}{c|}{\multirow{2}[2]{*}{Approaches}} & \multicolumn{2}{c|}{\FMALPACA} & \multicolumn{2}{c|}{\FMBENCH} & \multicolumn{2}{c|}{\FMALPACADF} & \multicolumn{2}{c}{\FMBENCHDF} \\
            &       &       & \multicolumn{1}{c}{FC} & \multicolumn{1}{c|}{WP} & \multicolumn{1}{c}{FC} & \multicolumn{1}{c|}{WP} & \multicolumn{1}{c}{FC} & \multicolumn{1}{c|}{WP} & \multicolumn{1}{c}{FC} & \multicolumn{1}{c}{WP} \\
        \midrule
        \multirow{9}[4]{*}{\begin{sideways}llama3.1-instruct\end{sideways}} & \multirow{6}[2]{*}{8B} & \ZeroShot & 7.86  & 1.43  & 8.33  & 0.00     & 11.79 & 1.07  & 8.33  & 1.67 \\
            &       & \FineTuneF & 24.29 & 1.07  & 51.67 & 0.00     & 50.36 & 2.14  & 65.00    & 0.00 \\
            &       & \FineTuneUF & 10.00    & 0.36  & 10.00    & 0.00     & 18.57 & 1.43  & 11.67 & 1.67 \\
            &       & \FineTuneTF & 13.57 & 3.57  & 3.33  & 0.00     & 26.07 & 1.07  & 20.00    & 1.67 \\
            &       & \LLMVer & 31.43 & 1.43  & 36.67 & 5.00     & 36.43 & 1.43  & 43.33 & 3.33 \\
            &       & \Solver (our) & \textbf{78.21} & \textbf{7.50} & \textbf{81.67} & \textbf{5.00} & \textbf{75.71} & \textbf{9.29} & \textbf{76.67} & \textbf{10.00} \\
    \cmidrule{2-11}          & \multirow{3}[2]{*}{70B} & \ZeroShot & 23.21 & 8.21  & 25.00    & 8.33  & 30.00    & 4.64  & 35.00    & 5.00 \\
            &       & \LLMVer & 62.50  & 29.29 & 63.33 & 33.33 & 62.86 & 21.79 & 78.33 & 18.33 \\
            &       & \Solver (our) & \textbf{78.93} & \textbf{32.86} & \textbf{85.00} & \textbf{36.67} & \textbf{84.29} & \textbf{29.64} & \textbf{91.67} & \textbf{28.33} \\
        \midrule
        \multirow{7}[4]{*}{\begin{sideways}qwen2.5-instruct\end{sideways}} & \multirow{4}[2]{*}{7B} & \ZeroShot & 22.50  & 3.57  & 21.67 & 5.00     & 31.79 & 2.50  & 41.67 & 0.00 \\
            &       & \FineTuneF & 36.43 & 7.14  & 31.67 & 0.00     & 52.86 & 3.57  & 45.00    & 0.00 \\
            &       & \LLMVer & 83.57 & 36.07 & 95.00    & 40.00    & 83.93 & \textbf{21.43} & 90.00    & 21.67 \\
            &       & \Solver (our) & \textbf{90.36} & \textbf{46.79} & \textbf{96.67} & \textbf{43.33} & \textbf{88.57} & 21.07 & \textbf{96.67} & \textbf{23.33} \\
    \cmidrule{2-11}          & \multirow{3}[2]{*}{72B} & \ZeroShot & 32.14 & 10.71 & 40.00    & 8.33  & 46.43 & 7.14  & 68.33 & 0.00 \\
            &       & \LLMVer & 72.86 & 8.57  & 70.00    & 1.67  & 74.64 & 5.71  & 78.33 & \textbf{5.00} \\
            &       & \Solver (our) & \textbf{85.00} & \textbf{13.21} & \textbf{86.67} & \textbf{10.00} & \textbf{88.21} & \textbf{8.21} & \textbf{90.00} & 3.33 \\
        \midrule
        \multirow{7}[4]{*}{\begin{sideways}\shortstack{qwen2.5\\-coder-instruct}\end{sideways}} & \multirow{4}[2]{*}{7B} & \ZeroShot & 2.86  & 0.71  & 5.00     & 0.00     & 2.50  & 0.36  & 0.00     & 0.00 \\
            &       & \FineTuneF & 15.36 & 2.86  & 5.00     & 0.00     & 30.00    & 2.86  & 16.67 & 0.00 \\
            &       & \LLMVer & 30.00    & 4.29  & 30.00    & 0.00     & 36.43 & 2.86  & 20.00    & 1.67 \\
            &       & \Solver (our) & \textbf{88.21} & \textbf{16.07} & \textbf{73.33} & \textbf{11.67} & \textbf{90.71} & \textbf{12.86} & \textbf{93.33} & \textbf{13.33} \\
    \cmidrule{2-11}          & \multirow{3}[2]{*}{32B} & \ZeroShot & 32.86 & 12.50 & 36.67 & 8.33  & 46.79 & 5.00     & 78.33 & 3.33 \\
            &       & \LLMVer & 66.07 & 18.57 & 76.67 & 21.67 & 58.21 & 10.36 & 66.67 & 11.67 \\
            &       & \Solver (our) & \textbf{77.14} & \textbf{29.29} & \textbf{85.00} & \textbf{36.67} & \textbf{76.43} & \textbf{18.57} & \textbf{81.67} & \textbf{38.33} \\
        \midrule
        \multirow{4}[2]{*}{\begin{sideways}\shortstack{deepseek\\-coder\\-instruct}\end{sideways}} & \multirow{4}[2]{*}{7B} & \ZeroShot & 19.64 & 1.07  & 6.67  & 3.33  & 19.64 & 1.07  & 6.67  & 3.33 \\
            &       & \FineTuneF & 23.93 & 0.00     & 38.33 & 0.00     & 59.29 & 3.57  & 48.33 & 0.00 \\
            &       & \LLMVer & 43.57 & 5.36  & 28.33 & 3.33  & 62.50  & 3.57  & 53.33 & 3.33 \\
            &       & \Solver (our) & \textbf{75.00} & \textbf{11.61} & \textbf{72.22} & \textbf{5.56} & \textbf{83.04} & \textbf{4.46} & \textbf{91.67} & \textbf{8.33} \\
        \midrule
        \multirow{3}[2]{*}{\begin{sideways}\shortstack{qwen3\\-AWQ}\end{sideways}} & \multirow{3}[2]{*}{32B} & \ZeroShot & 23.57 & 8.57  & 31.67 & 6.67  & 22.14 & 4.64  & 36.67 & 1.67 \\
            &       & \LLMVer & 92.14 & \textbf{57.86} & 93.33 & 66.67 & 88.57 & 46.43 & 96.67 & 56.67 \\
            &       & \Solver (our) & \textbf{92.5} & 56.79 & \textbf{100.00} & \textbf{75.00} & \textbf{91.43} & \textbf{55.71} & \textbf{100.00} & \textbf{58.33} \\
        \midrule
        \multirow{3}[2]{*}{\begin{sideways}\shortstack{deepseek\\-r1}\end{sideways}} & \multirow{3}[2]{*}{671B} & \ZeroShot & 44.64 & 15.00    & 55.00    & 20.00    & 28.21 & 5.36  & 36.67 & 1.67 \\
            &       & \LLMVer & 94.64 & 66.79 & 90.00    & 71.67 & 92.50 & 68.21 & 85.00    & 65.00 \\
            &       & \Solver (our) & \textbf{96.43} & \textbf{72.14} & \textbf{100.00} & \textbf{78.33} & \textbf{98.21} & \textbf{77.14} & \textbf{100.00} & \textbf{88.33} \\
        \midrule
        \multirow{3}[2]{*}{\begin{sideways}\shortstack{gemini\\-2.5-pro}\end{sideways}} & \multirow{3}[2]{*}{/} & \ZeroShot & 68.21 & 43.93 & 66.67 & 43.33 & 63.57 & 29.29 & 73.33 & 26.67 \\
            &       & \LLMVer & \textbf{98.93} & 85.36 & 98.33 & 80    & 96.79 & 81.79 & \textbf{100.00} & 81.67 \\
            &       & \Solver (our) & 98.21 & \textbf{90.36} & \textbf{100.00} & \textbf{86.67} & \textbf{99.29} & \textbf{88.21} & \textbf{100.00} & \textbf{83.33} \\
        \bottomrule
        \end{tabular}%
    }
\end{table}%

\begin{table}[htbp]
    \centering
    \caption{Pass@1 (\%) across different approaches.}\label{tab:result-pass1}%
    \scalebox{0.9}{
        \begin{tabular}{c|c|l|rr|rr|rr|rr}
        \toprule
        \multirow{2}[2]{*}{LLMs} & \multirow{2}[2]{*}{Sizes} & \multicolumn{1}{c|}{\multirow{2}[2]{*}{Approaches}} & \multicolumn{2}{c|}{\FMALPACA} & \multicolumn{2}{c|}{\FMBENCH} & \multicolumn{2}{c|}{\FMALPACADF} & \multicolumn{2}{c}{\FMBENCHDF} \\
            &       &       & \multicolumn{1}{c}{FC} & \multicolumn{1}{c|}{WP} & \multicolumn{1}{c}{FC} & \multicolumn{1}{c|}{WP} & \multicolumn{1}{c}{FC} & \multicolumn{1}{c|}{WP} & \multicolumn{1}{c}{FC} & \multicolumn{1}{c}{WP} \\
        \midrule
        \multirow{9}[4]{*}{\begin{sideways}llama3.1-instruct\end{sideways}} & \multirow{6}[2]{*}{8B} & \ZeroShot & 0.00     & 0.00     & 16.67 & 0.00     & 12.50  & 1.79  & 8.33  & 0.00 \\
            &       & \FineTuneF & 28.57 & 0.00     & 50.00    & 0.00     & 46.43 & 1.79  & 58.33 & 0.00 \\
            &       & \FineTuneUF & 12.50  & 0.00     & 0.00     & 0.00     & 12.50  & 0.00     & 8.33  & 0.00 \\
            &       & \FineTuneTF & 14.29 & 5.36  & 0.00     & 0.00     & 30.36 & 1.79  & 25.00    & 0.00 \\
            &       & \LLMVer & 32.14 & 1.79  & 41.67 & \textbf{16.67} & 41.07 & 1.79  & 50.00    & \textbf{8.33} \\
            &       & \Solver (our) & \textbf{80.36} & \textbf{8.93} & \textbf{83.33} & 8.33  & \textbf{73.21} & \textbf{7.14} & \textbf{66.67} & \textbf{8.33} \\
        \cmidrule{2-11}      & \multirow{3}[2]{*}{70B} & \ZeroShot & 30.36 & 8.93  & 25.00    & 0.00     & 37.50  & 5.36  & 33.33 & 0.00 \\
            &       & \LLMVer & 55.36 & 26.79 & 66.67 & \textbf{41.67} & 64.29 & 32.14 & 75.00    & \textbf{25.00} \\
            &       & \Solver (our) & \textbf{80.36} & \textbf{28.57} & \textbf{83.33} & 25.00    & \textbf{83.93} & \textbf{41.07} & \textbf{100.00} & 16.67 \\
        \midrule
        \multirow{7}[4]{*}{\begin{sideways}qwen2.5-instruct\end{sideways}} & \multirow{4}[2]{*}{7B} & \ZeroShot & 21.43 & 3.57  & 25.00    & 8.33  & 33.93 & 3.57  & 50.00    & 0.00 \\
            &       & \FineTuneF & 35.71 & 3.57  & 33.33 & 0.00     & 51.79 & 3.57  & 50.00    & 0.00 \\
            &       & \LLMVer & 80.36 & 41.07 & 91.67 & \textbf{50.00} & 89.29 & \textbf{21.43} & 83.33 & \textbf{16.67} \\
            &       & \Solver (our) & \textbf{92.86} & \textbf{44.64} & \textbf{100.00} & 33.33 & \textbf{91.07} & \textbf{21.43} & \textbf{91.67} & \textbf{16.67} \\
        \cmidrule{2-11}      & \multirow{3}[2]{*}{72B} & \ZeroShot & 25.00    & 7.14  & 33.33 & 8.33  & 46.43 & \textbf{7.14} & 66.67 & \textbf{0.00} \\
            &       & \LLMVer & 71.43 & 10.71 & 58.33 & 0.00     & 71.43 & 1.79  & 75.00    & \textbf{0.00} \\
            &       & \Solver (our) & \textbf{83.93} & \textbf{12.50} & \textbf{100.00} & \textbf{16.67} & \textbf{91.07} & \textbf{7.14} & \textbf{83.33} & \textbf{0.00} \\
        \midrule
        \multirow{7}[4]{*}{\begin{sideways}\shortstack{qwen2.5\\-coder-instruct}\end{sideways}} & \multirow{4}[2]{*}{7B} & \ZeroShot & 0.00     & 0.00     & 16.67 & 0.00     & 1.79  & 1.79  & 0.00     & 0.00 \\
            &       & \FineTuneF & 14.29 & 3.57  & 8.33  & 0.00     & 35.71 & 5.36  & 16.67 & 0.00 \\
            &       & \LLMVer & 35.71 & 0.00     & 33.33 & 0.00     & 30.36 & 0.00     & 16.67 & 0.00 \\
            &       & \Solver (our) & \textbf{92.86} & \textbf{17.86} & \textbf{75.00} & \textbf{8.33} & \textbf{96.43} & \textbf{12.50} & \textbf{83.33} & \textbf{8.33} \\
        \cmidrule{2-11}      & \multirow{3}[2]{*}{32B} & \ZeroShot & 30.36 & 10.71 & 33.33 & 8.33  & 44.64 & 3.57  & 75.00    & 0.00 \\
            &       & \LLMVer & 60.71 & 21.43 & 75.00    & 16.67 & 62.50  & 8.93  & 58.33 & 8.33 \\
            &       & \Solver (our) & \textbf{75.00} & \textbf{28.57} & \textbf{83.33} & \textbf{33.33} & \textbf{76.79} & \textbf{16.07} & \textbf{66.67} & \textbf{33.33} \\
        \midrule
        \multirow{4}[2]{*}{\begin{sideways}\shortstack{deepseek\\-coder\\-instruct}\end{sideways}} & \multirow{4}[2]{*}{7B} & \ZeroShot & 21.43 & 0.00     & 0.00     & 0.00     & 21.43 & 0.00     & 8.33  & 0.00 \\
            &       & \FineTuneF & 25.00    & 0.00     & 33.33 & 0.00     & 58.93 & \textbf{3.57} & 41.67 & 0.00 \\
            &       & \LLMVer & 35.71 & 3.57  & 33.33 & 0.00     & 62.50  & \textbf{3.57} & 58.33 & 0.00 \\
            &       & \Solver (our) & \textbf{67.86} & \textbf{8.93} & \textbf{75.00} & \textbf{8.33} & \textbf{80.36} & \textbf{3.57} & \textbf{100.00} & \textbf{8.33} \\
        \midrule
        \multirow{3}[2]{*}{\begin{sideways}\shortstack{qwen3\\-AWQ}\end{sideways}} & \multirow{3}[2]{*}{32B} & \ZeroShot & 17.86 & 3.57  & 33.33 & 0.00     & 25.00    & 5.36  & 50.00    & 0.00 \\
            &       & \LLMVer & \textbf{96.43} & 51.79 & 91.67 & 58.33 & \textbf{92.86} & 37.50  & \textbf{100.00} & 58.33 \\
            &       & \Solver (our) & 92.86 & \textbf{57.14} & \textbf{100.00} & \textbf{75.00} & 91.07 & \textbf{50.00} & \textbf{100.00} & \textbf{66.67} \\
        \midrule
        \multirow{3}[2]{*}{\begin{sideways}\shortstack{deepseek\\-r1}\end{sideways}} & \multirow{3}[2]{*}{671B} & \ZeroShot & 44.64 & 17.86 & 66.67 & 16.67 & 32.14 & 12.50  & 50.00    & 8.33 \\
            &       & \LLMVer & 96.43 & 67.86 & \textbf{100.00} & \textbf{75.00} & 96.43 & 71.43 & 91.67 & 75.00 \\
            &       & \Solver (our) & \textbf{98.21} & \textbf{76.79} & \textbf{100.00} & \textbf{75.00} & \textbf{98.21} & \textbf{78.57} & \textbf{100.00} & \textbf{83.33} \\
        \midrule
        \multirow{3}[2]{*}{\begin{sideways}\shortstack{gemini\\-2.5-pro}\end{sideways}} & \multirow{3}[2]{*}{/} & \ZeroShot & 67.86 & 41.07 & 50.00    & 41.67 & 71.43 & 26.79 & 75.00    & \textbf{33.33} \\
            &       & \LLMVer & \textbf{100.00} & 82.14 & \textbf{100.00} & \textbf{83.33} & \textbf{98.21} & 78.57 & \textbf{100.00} & \textbf{83.33} \\
            &       & \Solver (our) & \textbf{100.00} & \textbf{96.43} & \textbf{100.00} & \textbf{83.33} & \textbf{98.21} & \textbf{89.29} & \textbf{100.00} & \textbf{83.33} \\
        \bottomrule
        \end{tabular}%
    }
\end{table}%

\begin{table}[htbp]
    \centering
    \caption{Pass@5 (\%) across different approaches.}\label{tab:result-pass5}%
    \scalebox{0.9}{
        \begin{tabular}{c|c|l|rr|rr|rr|rr}
        \toprule
        \multirow{2}[2]{*}{LLMs} & \multirow{2}[2]{*}{Sizes} & \multicolumn{1}{c|}{\multirow{2}[2]{*}{Approaches}} & \multicolumn{2}{c|}{\FMALPACA} & \multicolumn{2}{c|}{\FMBENCH} & \multicolumn{2}{c|}{\FMALPACADF} & \multicolumn{2}{c}{\FMBENCHDF} \\
            &       &       & \multicolumn{1}{c}{FC} & \multicolumn{1}{c|}{WP} & \multicolumn{1}{c}{FC} & \multicolumn{1}{c|}{WP} & \multicolumn{1}{c}{FC} & \multicolumn{1}{c|}{WP} & \multicolumn{1}{c}{FC} & \multicolumn{1}{c}{WP} \\
        \midrule
        \multirow{9}[4]{*}{\begin{sideways}llama3.1-instruct\end{sideways}} & \multirow{6}[2]{*}{8B} & \ZeroShot & 33.93 & 5.36  & 33.33 & 0.00     & 42.86 & 5.36  & 41.67 & 8.33 \\
            &       & \FineTuneF & 48.21 & 5.36  & 66.67 & 0.00     & 67.86 & 5.36  & 75.00    & 0.00 \\
            &       & \FineTuneUF & 28.57 & 1.79  & 25.00    & 0.00     & 39.29 & 5.36  & 33.33 & 8.33 \\
            &       & \FineTuneTF & 26.79 & 8.93  & 16.67 & 0.00     & 37.50  & 3.57  & 41.67 & 8.33 \\
            &       & \LLMVer & 78.57 & 7.14  & \textbf{100.00} & \textbf{16.67} & 82.14 & 5.36  & 83.33 & 8.33 \\
            &       & \Solver (our) & \textbf{96.43} & \textbf{17.86} & \textbf{100.00} & \textbf{16.67} & \textbf{94.64} & \textbf{17.86} & \textbf{100.00} & \textbf{16.67} \\
        \cmidrule{2-11}      & \multirow{3}[2]{*}{70B} & \ZeroShot & 51.79 & 21.43 & 66.67 & 33.33 & 62.50  & 10.71 & 91.67 & 16.67 \\
            &       & \LLMVer & 89.29 & 51.79 & \textbf{100.00} & 66.67 & \textbf{91.07} & 46.43 & 91.67 & 25.00 \\
            &       & \Solver (our) & \textbf{91.07} & \textbf{55.36} & \textbf{100.00} & \textbf{75.00} & \textbf{91.07} & \textbf{53.57} & \textbf{100.00} & \textbf{50.00} \\
        \midrule
        \multirow{7}[4]{*}{\begin{sideways}qwen2.5-instruct\end{sideways}} & \multirow{4}[2]{*}{7B} & \ZeroShot & 53.57 & 12.50  & 41.67 & 8.33  & 62.50  & 8.93  & 83.33 & 0.00 \\
            &       & \FineTuneF & 57.14 & 12.50  & 50.00    & 0.00     & 67.86 & 7.14  & 75.00    & 0.00 \\
            &       & \LLMVer & 96.43 & 57.14 & \textbf{100.00} & \textbf{66.67} & 94.64 & \textbf{39.29} & \textbf{100.00} & \textbf{50.00} \\
            &       & \Solver (our) & \textbf{100.00} & \textbf{64.29} & \textbf{100.00} & \textbf{66.67} & \textbf{98.21} & 35.71 & \textbf{100.00} & 33.33 \\
        \cmidrule{2-11}      & \multirow{3}[2]{*}{72B} & \ZeroShot & 64.29 & 21.43 & 66.67 & 16.67 & 76.79 & 14.29 & 91.67 & 0.00 \\
            &       & \LLMVer & 91.07 & 25.00    & \textbf{100.00} & 8.33  & \textbf{98.21} & 17.86 & \textbf{100.00} & \textbf{25.00} \\
            &       & \Solver (our) & \textbf{96.43} & \textbf{28.57} & \textbf{100.00} & \textbf{33.33} & \textbf{98.21} & \textbf{28.57} & \textbf{100.00} & 16.67 \\
        \midrule
        \multirow{7}[4]{*}{\begin{sideways}\shortstack{qwen2.5\\-coder-instruct}\end{sideways}} & \multirow{4}[2]{*}{7B} & \ZeroShot & 10.71 & 1.79  & 16.67 & 0.00     & 8.93  & 1.79  & 0.00     & 0.00 \\
            &       & \FineTuneF & 26.79 & 8.93  & 8.33  & 0.00     & 48.21 & 8.93  & 25.00    & 0.00 \\
            &       & \LLMVer & 76.79 & 14.29 & 83.33 & 0.00     & 82.14 & 8.93  & 83.33 & 8.33 \\
            &       & \Solver (our) & \textbf{100.00} & \textbf{25.00} & \textbf{100.00} & \textbf{25.00} & \textbf{100.00} & \textbf{32.14} & \textbf{100.00} & \textbf{25.00} \\
        \cmidrule{2-11}      & \multirow{3}[2]{*}{32B} & \ZeroShot & 41.07 & 14.29 & 50.00    & 8.33  & 51.79 & 7.14  & 83.33 & 8.33 \\
            &       & \LLMVer & \textbf{82.14} & 26.79 & 83.33 & 33.33 & 78.57 & 21.43 & \textbf{91.67} & 33.33 \\
            &       & \Solver (our) & 80.36 & \textbf{37.5} & \textbf{91.67} & \textbf{41.67} & \textbf{82.14} & \textbf{33.93} & \textbf{91.67} & \textbf{50.00} \\
        \midrule
        \multirow{4}[2]{*}{\begin{sideways}\shortstack{deepseek\\-coder\\-instruct}\end{sideways}} & \multirow{4}[2]{*}{7B} & \ZeroShot & 32.14 & 3.57  & 25.00    & 8.33  & 32.14 & 3.57  & 16.67 & 8.33 \\
            &       & \FineTuneF & 30.36 & 0.00     & 50.00    & 0.00     & 60.71 & 3.57  & 66.67 & 0.00 \\
            &       & \LLMVer & 85.71 & 14.29 & 75.00    & \textbf{16.67} & 94.64 & \textbf{10.71} & 83.33 & 16.67 \\
            &       & \Solver (our) & \textbf{91.07} & \textbf{17.86} & \textbf{91.67} & \textbf{16.67} & \textbf{96.43} & 7.14  & \textbf{100.00} & \textbf{25.00} \\
        \midrule
        \multirow{3}[2]{*}{\begin{sideways}\shortstack{qwen3\\-AWQ}\end{sideways}} & \multirow{3}[2]{*}{32B} & \ZeroShot & 50.00    & 19.64 & 66.67 & 16.67 & 55.36 & 12.50  & 91.67 & 8.33 \\
            &       & \LLMVer & \textbf{100.00} & 80.36 & \textbf{100.00} & \textbf{91.67} & \textbf{100.00} & 76.79 & \textbf{100.00} & \textbf{91.67} \\
            &       & \Solver (our) & \textbf{100.00} & \textbf{83.93} & \textbf{100.00} & 83.33 & \textbf{100.00} & \textbf{82.14} & \textbf{100.00} & 75.00 \\
        \midrule
        \multirow{3}[2]{*}{\begin{sideways}\shortstack{deepseek\\-r1}\end{sideways}} & \multirow{3}[2]{*}{671B} & \ZeroShot & 64.29 & 30.36 & 66.67 & 50.00    & 60.71 & 19.64 & 66.67 & 8.33 \\
            &       & \LLMVer & \textbf{100.00} & 85.71 & \textbf{100.00} & \textbf{91.67} & \textbf{100.00} & 85.71 & \textbf{100.00} & 91.67 \\
            &       & \Solver (our) & \textbf{100.00} & \textbf{87.5} & \textbf{100.00} & \textbf{91.67} & \textbf{100.00} & \textbf{91.07} & \textbf{100.00} & \textbf{100.00} \\
        \midrule
        \multirow{3}[2]{*}{\begin{sideways}\shortstack{gemini\\-2.5-pro}\end{sideways}} & \multirow{3}[2]{*}{/} & \ZeroShot & 89.29 & 69.64 & 91.67 & 58.33 & 91.07 & 50.00    & 91.67 & 50.00 \\
            &       & \LLMVer & \textbf{100.00} & \textbf{98.21} & \textbf{100.00} & \textbf{91.67} & \textbf{100.00} & 91.07 & \textbf{100.00} & \textbf{91.67} \\
            &       & \Solver (our) & \textbf{100.00} & \textbf{98.21} & \textbf{100.00} & \textbf{91.67} & \textbf{100.00} & \textbf{94.64} & \textbf{100.00} & 83.33 \\
        \bottomrule
    \end{tabular}%
    }
\end{table}%

\begin{table}[t]
    \centering
    \caption{Average number of LLM calls across \LLMVer and \Solver, where
    `DSCI' is short for `deepseek-coder-instruct', 
    `QWA' is short for `qwen3-AWQ', 
    `DSR' is short for `deepseek-R1', and
    `GMP' is short for `gemini-2.5-pro'.
    The smaller, the better.}\label{tab:result-num-LLM}%
    \scalebox{0.9}{
        \begin{tabular}{c|c|l|rrrr}
        \toprule
        \multicolumn{1}{c|}{LLMs}  & \multicolumn{1}{c|}{Sizes} & \multicolumn{1}{c|}{Approaches} & \multicolumn{1}{c}{\FMALPACA} & \multicolumn{1}{c}{\FMBENCH} & \multicolumn{1}{c}{\FMALPACADF} & \multicolumn{1}{c}{\FMBENCHDF} \\
        \midrule
        \multirow{4}[4]{*}{\begin{sideways}\shortstack{llama3.1\\-instruct}\end{sideways}} & \multirow{2}[2]{*}{8B} & \LLMVer & 39.57 & 39.02 & 39.46 & 38.77 \\
            &       & \Solver (our) & \textbf{38.11} & \textbf{38.40} & \textbf{38.10} & \textbf{36.52} \\
    \cmidrule{2-7}          & \multirow{2}[2]{*}{70B} & \LLMVer & 30.49 & 30.80  & 32.82 & 34.45 \\
            &       & \Solver (our) & \textbf{29.41} & \textbf{28.97} & \textbf{31.02} & \textbf{31.72} \\
        \midrule
        \multirow{4}[4]{*}{\begin{sideways}\shortstack{qwen2.5\\-instruct}\end{sideways}} & \multirow{2}[2]{*}{7B} & \LLMVer & 29.66 & \textbf{28.77} & 34.15 & \textbf{34.02} \\
            &       & \Solver (our) & \textbf{27.33} & 29.02 & \textbf{33.96} & 35.33 \\
    \cmidrule{2-7}          & \multirow{2}[2]{*}{72B} & \LLMVer & 37.39 & 39.47 & 38.22 & \textbf{38.42} \\
            &       & \Solver (our) & \textbf{36.26} & \textbf{37.10} & \textbf{37.71} & 39.75 \\
        \midrule
        \multirow{4}[4]{*}{\begin{sideways}\shortstack{qwen2.5\\-coder\\-instruct}\end{sideways}} & \multirow{2}[2]{*}{7B} & \LLMVer & 38.94 & 40.00    & 39.33 & 39.87 \\
            &       & \Solver (our) & \textbf{36.04} & \textbf{37.85} & \textbf{36.89} & \textbf{36.32} \\
    \cmidrule{2-7}          & \multirow{2}[2]{*}{32B} & \LLMVer & 33.10  & 31.77 & 36.15 & 35.57 \\
            &       & \Solver (our) & \textbf{30.18} & \textbf{26.78} & \textbf{34.42} & \textbf{27.58} \\
        \midrule
        \multirow{2}[2]{*}{\begin{sideways}\shortstack{DSCI}\end{sideways}} & \multirow{2}[2]{*}{7B} & \LLMVer & 38.76 & 39.105 & 38.73 & 39.1 \\
            &       & \Solver (our) & \textbf{37.83} & \textbf{38.89} & \textbf{38.47} & \textbf{38.22} \\
        \midrule
        \multirow{2}[2]{*}{\begin{sideways}\shortstack{QWA}\end{sideways}} & \multirow{2}[2]{*}{32B} & \LLMVer & \textbf{23.30} & 19.50  & 27.43 & 24.38 \\
            &       & \Solver (our) & 23.36 & \textbf{19.22} & \textbf{25.71} & \textbf{23.25} \\
        \midrule
        \multirow{2}[2]{*}{\begin{sideways}\shortstack{DSR}\end{sideways}} & \multirow{2}[2]{*}{671B} & \LLMVer & \textbf{17.54} & 17.30  & 17.54 & 16.60 \\
            &       & \Solver (our) & 17.76 & \textbf{16.30} & \textbf{17.36} & \textbf{12.73} \\
        \midrule
        \multirow{2}[2]{*}{\begin{sideways}\shortstack{GMP}\end{sideways}} & \multirow{2}[2]{*}{/} & \LLMVer & 10.82 & 10.52 & 11.41 & 10.27 \\
            &       & \Solver (our) & \textbf{9.85} & \textbf{8.98} & \textbf{10.39} & \textbf{9.90} \\
        \bottomrule
        \end{tabular}%
    }
\end{table}%

\begin{table}[t]
  \centering
  \caption{Ablation results (\%) about diverse individual initialization, collaborative crossover, and self-reflective mutation, where 
        `ave. Pass@1' means average Pass@1,
        `w/o RS' indicates \Solver without reasoning strategies in the individual initialization, \ie, all individuals are initialized using the same prompt, 
        `w/o SBP' indicates \Solver without SBP in the individual initialization, \ie, all individuals are initialized using CoT, 
        `w/o CoT' indicates \Solver without CoT in the individual initialization, \ie, all individuals are initialized using SBP,
        `w/o E' indicates \Solver without the evolutionary process, 
        `w/o E \& TS' indicates \Solver without the evolutionary process and directly requires LLMs to synthesize ACSL-annotated code from requirements, 
        `w/o crossover' indicates \Solver without the collaborative crossover in the evolutionary process, and
        `w/o mutation' indicates \Solver without the mutation in the evolutionary process.}\label{tab:result-ablation}%
    \scalebox{0.9}{
        \begin{tabular}{l|rrr|rrr}
        \toprule
        \multicolumn{1}{c|}{\multirow{2}[2]{*}{Approaches}} & \multicolumn{3}{c|}{FC} & \multicolumn{3}{c}{WP} \\
            & \multicolumn{1}{c}{ave. Pass@1} & \multicolumn{1}{c}{Pass@1} & \multicolumn{1}{c|}{Pass@5} & \multicolumn{1}{c}{ave. Pass@1} & \multicolumn{1}{c}{Pass@1} & \multicolumn{1}{c}{Pass@5} \\
        \midrule
        \Solver & \textbf{100.00} & \textbf{100.00} & \textbf{100.00} & \textbf{75.00} & \textbf{83.33} & \textbf{83.33} \\
        \midrule
        w/o RS & 96.67 & 91.67 & \textbf{100.00} & 60.00    & 58.33 & \textbf{83.33} \\
        w/o SBP & 95.00    & 83.33 & \textbf{100.00} & 51.67 & 50.00    & 75.00 \\
        w/o CoT & 96.67 & 91.67 & \textbf{100.00} & 53.33 & 66.67 & \textbf{83.33} \\
        \midrule
        w/o E & 71.67 & 91.67 & \textbf{100.00} & 25.00    & 25.00    & 50.00 \\
        w/o E \& TS & 55.00    & 58.33 & \textbf{100.00} & 10.00    & 8.33  & 16.67 \\
        \midrule
        w/o crossover & 91.70  & 91.67 & \textbf{100.00} & 51.67 & 50.00    & 75.00 \\
        w/o mutation & \textbf{100.00} & \textbf{100.00} & \textbf{100.00} & 61.67 & 58.33 & 75.00 \\
        \bottomrule
        \end{tabular}%
    }
\end{table}%

\subsubsection{Compared with SOTA Approaches.}

Table~\ref{tab:result-ave-pass1} summarizes the average Pass@1.
Overall, \Solver almost outperforms all approaches across all LLM backbones and datasets: average improvement of at least $17.64$\% FC and $6.29$\% WP.  
For example, on qwen2.5-coder-instruct (7B), \Solver achieves $88.21$\% FC and $16.07$\% WP on \FMALPACA, significantly surpassing the SOTA approach (\LLMVer: $30.00$\% FC and $4.29$\% WP). 
Similarly, on the LLMs with larger parameters, \eg, qwen2.5-instruct (72B) and deepseek-r1 (671B), \Solver maintains clear superiority in both verifiers. 
Table~\ref{tab:result-pass1} and Table~\ref{tab:result-pass5} summarizes the Pass@1 and Pass@5, respectively.
They exhibit a similar trend to the average Pass@1 (Table~\ref{tab:result-ave-pass1}). 

On the developer-friendly datasets, the advantage of \Solver becomes even more pronounced.
For all LLM backbones, compared with the SOTA approach (\LLMVer), \Solver achieves an average improvement of $17.42$\% FC and $5.85$\% WP on the original dataset, and an average improvement of $17.87$\% and $6.74$\% WP on the developer-friendly dataset.
For example, on qwen2.5-coder-instruct (32B), \Solver achieves $76.43$\% FC and $18.57$\% WP on \FMALPACADF, compared to the performance of \LLMVer $58.21$\% FC and $10.36$\% WP. 
This trend holds across most model families.
Besides, both the maximum $73.33$\% FC improvement and $26.66$\% WP improvement are achieved on developer-friendly datasets.
It demonstrates the strong capability of \Solver in inferring implicit formal specifications from purely functional descriptions.

When paired with SOTA LLMs such as gemini-2.5-Pro, \Solver achieves excellent pass rates on multiple datasets. 
Note that using these LLMs, which are claimed to possess powerful reasoning abilities, alone is still insufficient to generate high-quality verifiable code, shown the results of \ZeroShot.
This indicates that \Solver effectively leverages the enhanced reasoning and code generation capabilities of advanced LLMs, translating their raw potential into higher rates of formally verified code synthesis. 
At the same time, we also see how the improved performance of LLMs has benefited \Solver.
Therefore, the synergy between a powerful LLM backbone and our structured evolutionary search enables more robust exploration of the solution space. 
 
Despite leveraging domain-specific fine-tuning, the fine-tuning approaches demonstrate only modest improvements over zero-shot prompting and fall considerably short of \Solver on all datasets. 
These results suggest that fine-tuning does not adequately resolve the challenges of autoformalization, such as hallucination of syntax and semantic misalignment. 
This underscores the need for more sophisticated synthesis mechanisms beyond supervised adaptation.

Table~\ref{tab:result-num-LLM} compares the average number of LLM calls per instance between \LLMVer and \Solver. 
We can observe that \Solver requires fewer or comparable LLM calls in most cases. 
This indicates that \Solver not only enhances synthesis quality but also maintains computational efficiency.

Overall, these results confirm that \Solver effectively mitigates error propagation and hallucination issues inherent in single-agent iterative approaches, while successfully uncovering implicit knowledge necessary for formal verification.
Above conclusions answer RQ~\ref{que:experiment-1} and~\ref{que:experiment-2}.

\subsubsection{Ablation Study.}

The experiments are performed on \FMBENCH using qwen3-AWQ (32B), with results presented in Table~\ref{tab:result-ablation}. 
The results are similar on other datasets and with other LLMs.
When examining initialization reasoning strategies, `w/o RS' reduces average Pass@1 from $100.00$\% to $96.67$\% for FC and from $75.00$\% to $60.00$\% for WP. 
`w/o SBP' causes significant performance drops, especially in WP ($51.67$\% average Pass@1). 
While `w/o CoT' also degrades performance, though less severely than `w/o SBP'. 
This demonstrates the importance of reasoning diversity.
The evolutionary process emerges as a critical component, with its complete removal (`w/o E') reducing average Pass@1 to $71.67$\% on FC and causing a severe decline to $25$\% on WP. 
However, `w/o E \& TS' leads to a more pronounced degradation, underscoring the significance of the two-phase strategy.
For evolutionary operators, both `w/o crossover' and `w/o mutation' yield performance degradation. Notably, the impact is greater when crossover is removed. 
It indicates that although mutation and crossover jointly contribute to search-space exploration, crossover plays a dominant role in \Solver.

In summary, the ablation results collectively demonstrate that the complete \Solver framework is essential for achieving state-of-the-art performance in automated, verifiable code synthesis.
Above conclusions answer RQ~\ref{que:experiment-3}.

\subsubsection{Robustness to Hyperparameters.}

\begin{figure}[t]
    \centering
    \includegraphics[width=0.49\textwidth]{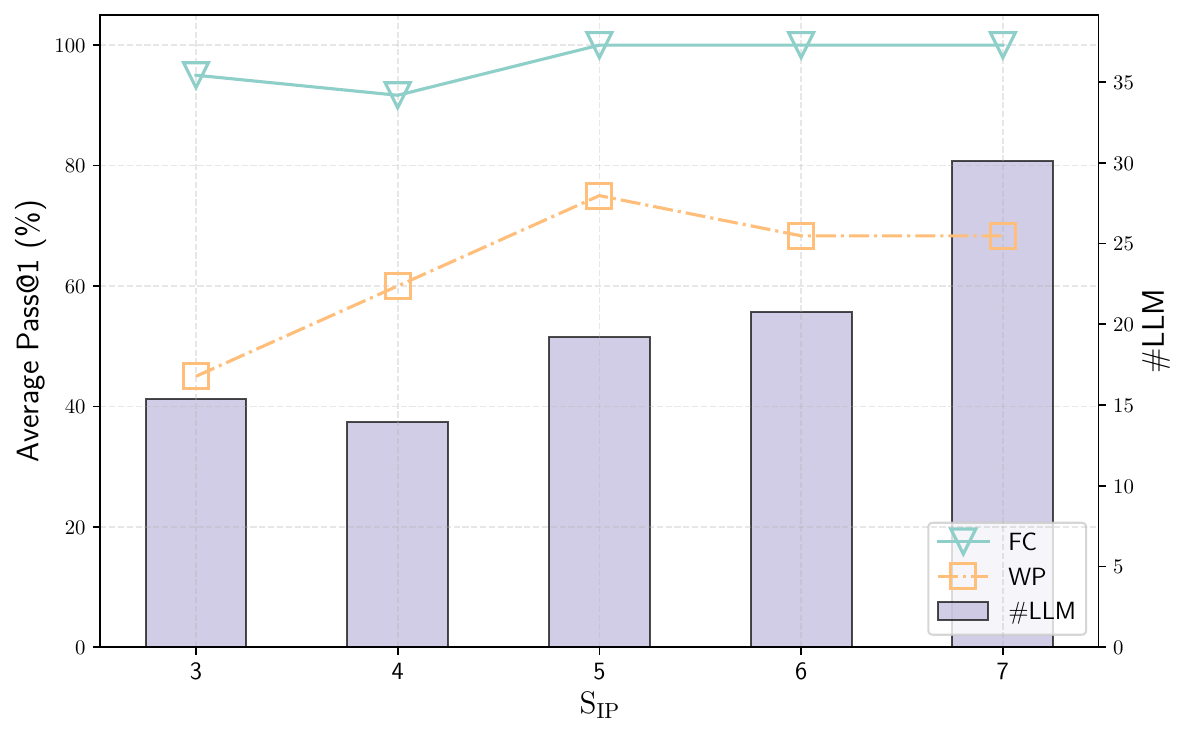}
    \includegraphics[width=0.49\textwidth]{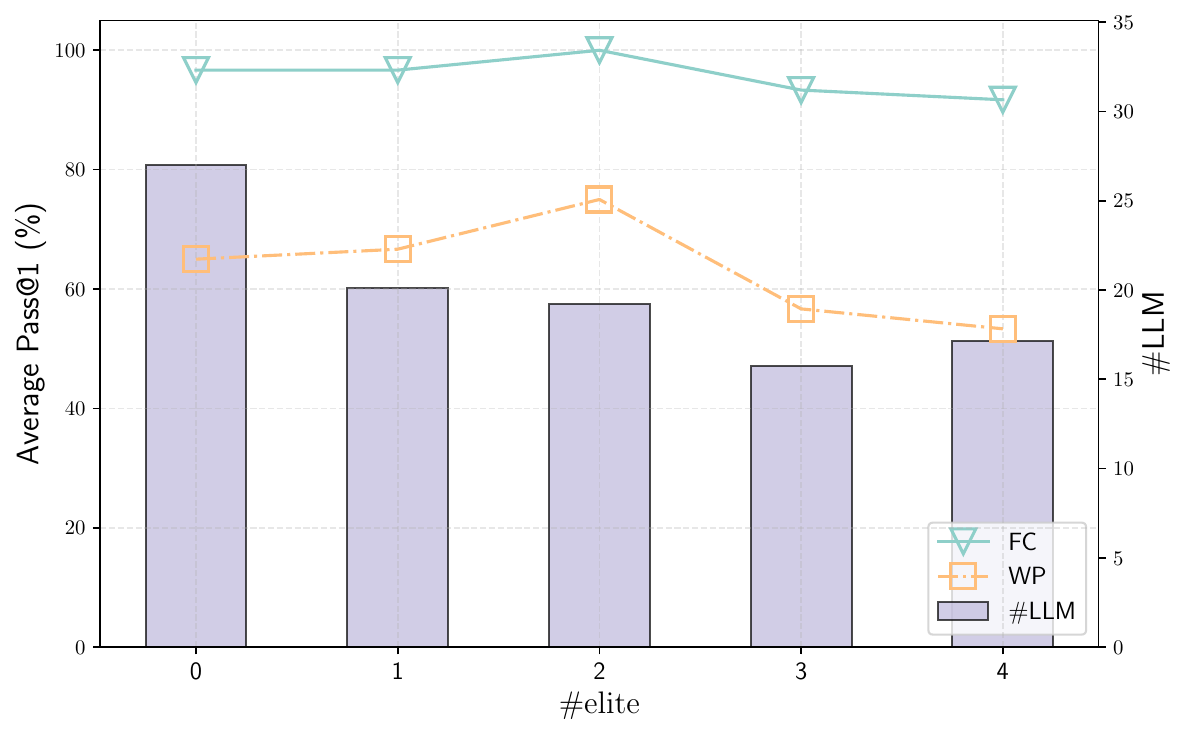}
    \includegraphics[width=0.49\textwidth]{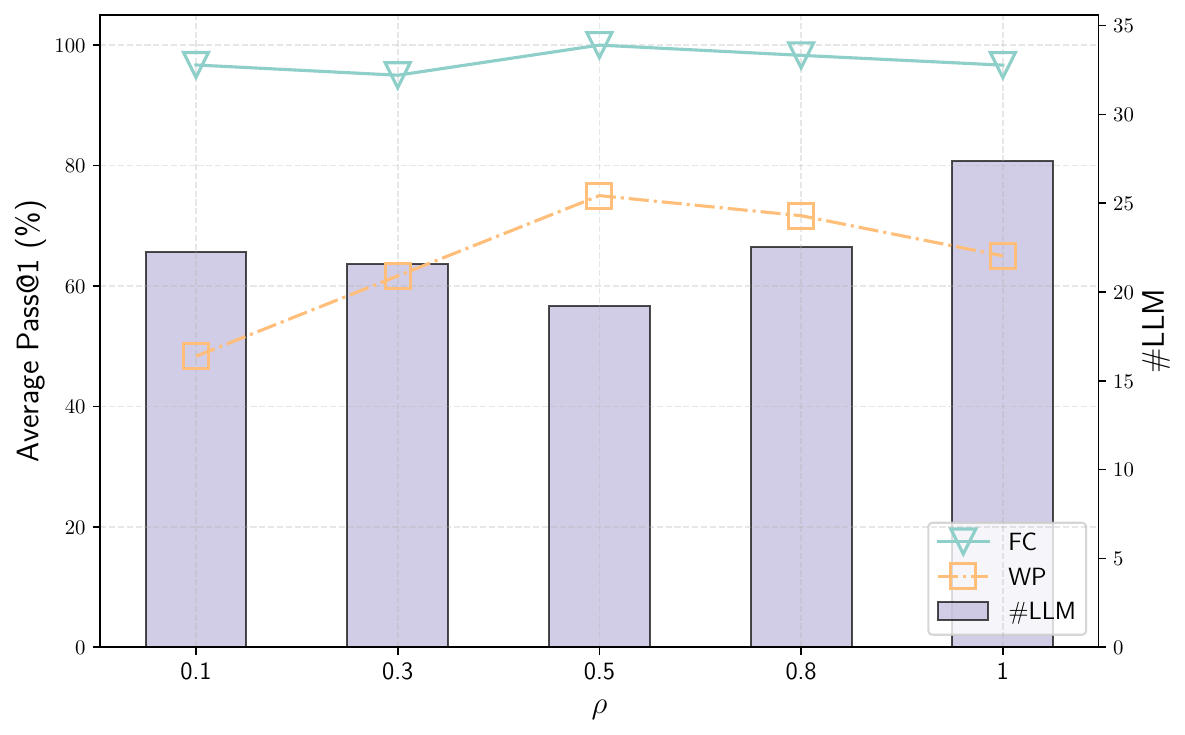}
    \includegraphics[width=0.49\textwidth]{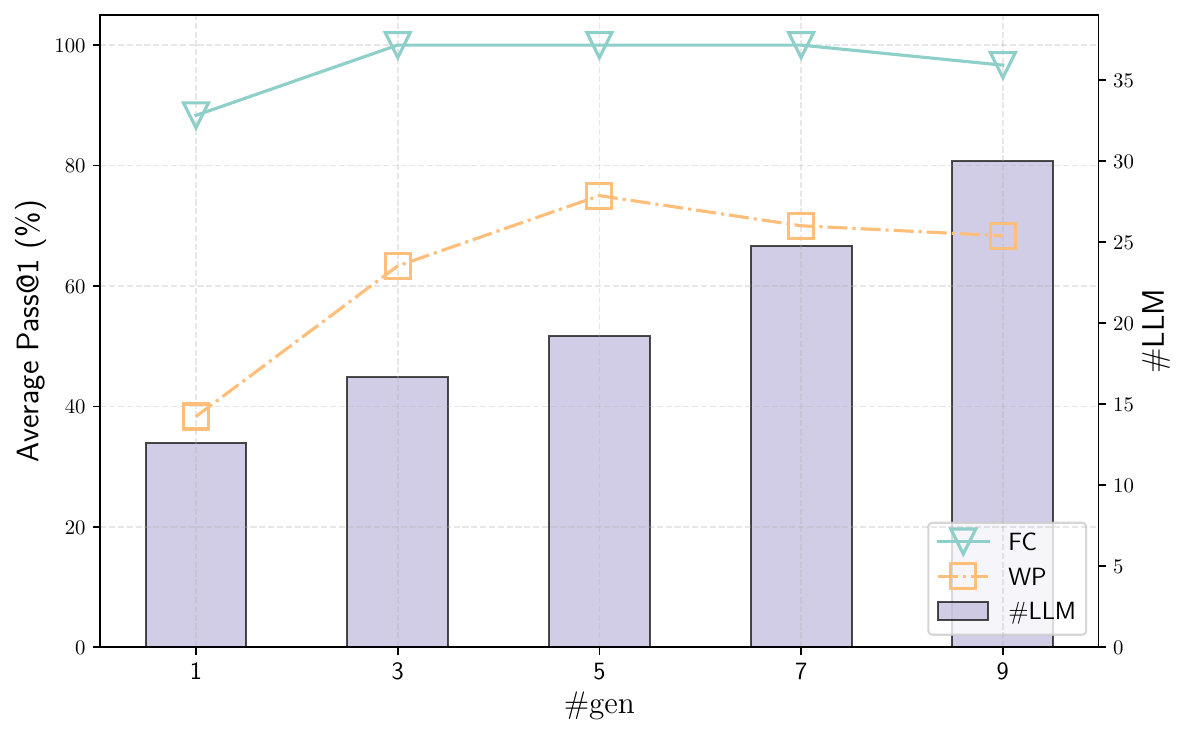}

    \caption{Comparisons for different settings of $\SizeIniPop$, $\NumElite$, $\MutateR$, and $\MaxGen$ over \FMBENCH.}
    \label{fig:result-hyperparameter-analysis}
\end{figure}

The main hyperparameters of \Solver are $\SizeIniPop$, $\NumElite$, $\MutateR$, and $\MaxGen$.
We conducted experiments to show the impact of different values of these hyperparameters, as illustrated in Fig.~\ref{fig:result-hyperparameter-analysis}.
Since the trends of all datasets are exactly the same, we only show the results for on \FMBENCH using qwen3-AWQ (32B).
We only vary the hyperparameter being discussed and keep the default values for the other hyperparameters.

As $\SizeIniPop$ increases, performance initially improves, then declines. 
Considering that increasing $\SizeIniPop$ will increase computational cost, an intermediate value of $\SizeIniPop$ likely yields the best performance, achieving near-optimal performance while balancing computational efficiency.
$\NumElite$ exhibits robust behavior, with average Pass@1 remaining stable for values ranging from $0$ to $2$, but slightly declines when $\NumElite = 3$. 
It is because increasing $\NumElite$, while guaranteeing high-quality individuals, can inhibit exploration, causing \Solver to easily get stuck in a local optimum.
It indicates that our default setting of $\NumElite = 2$ effectively preserves optimal solutions without sacrificing population diversity. 
$\MutateR$ and $\NumElite$ exhibit similar trends, with our default value of $\MutateR = 0.5$ providing an excellent balance between exploration and exploitation. 
For $\MaxGen$, we can find that excessive generations may lead to diminishing performance and increased computational cost. 
So the setting of $\MaxGen = 5$ achieves satisfactory results while maintaining reasonable computational overhead. 

In summary, increasing the values of the four hyperparameters has a positive impact on performance, but this positive effect gradually diminishes and may even become negative.
Besides, within a certain range, \Solver is robust to hyperparameters, which answers RQ~\ref{que:experiment-4}.

\subsection{Case Study}\label{sec:case-study}

\begin{figure}[t]
    \centering
    \begin{minipage}[t]{0.49\textwidth}
        \centering
        \includegraphics[width=6.2cm,height=4cm]{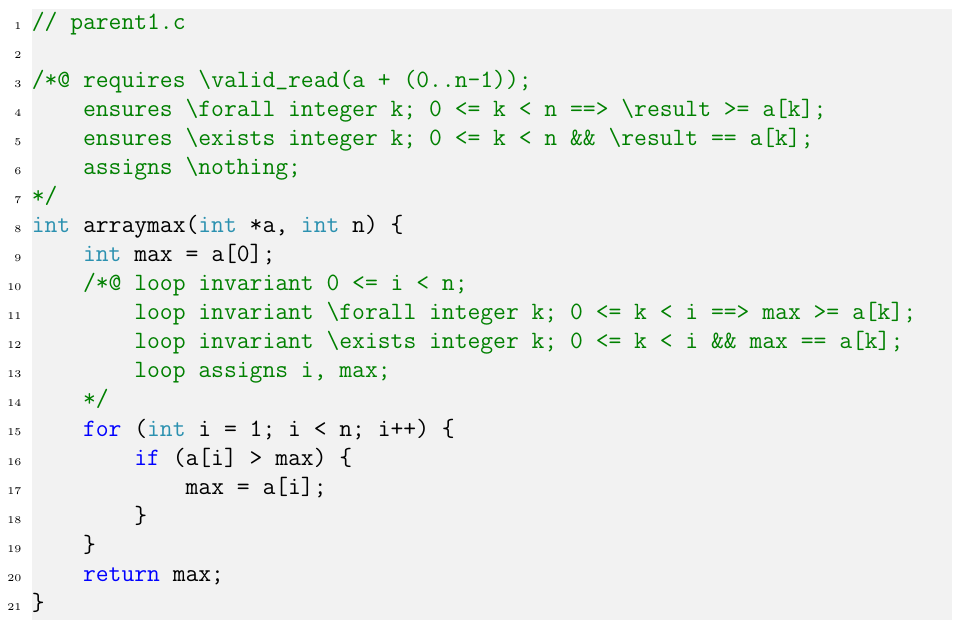}
    \end{minipage}
    \hfill
    \begin{minipage}[t]{0.49\textwidth}
        \centering
        \includegraphics[width=6.2cm,height=4cm]{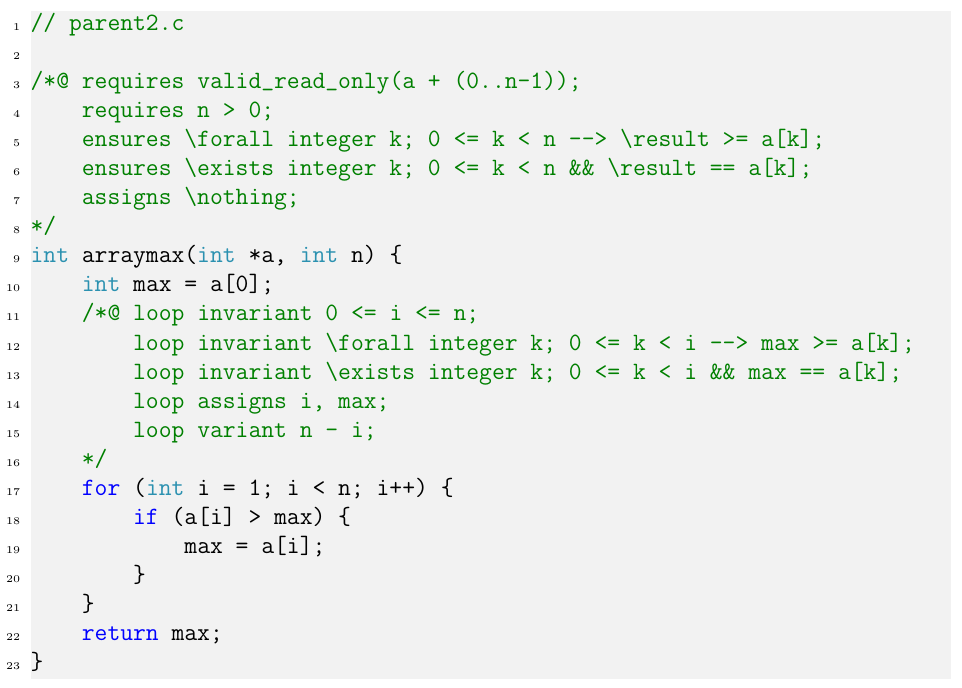}
    \end{minipage}


    \begin{minipage}[t]{0.49\textwidth}
        \centering
        \includegraphics[width=6.2cm,height=4.4cm]{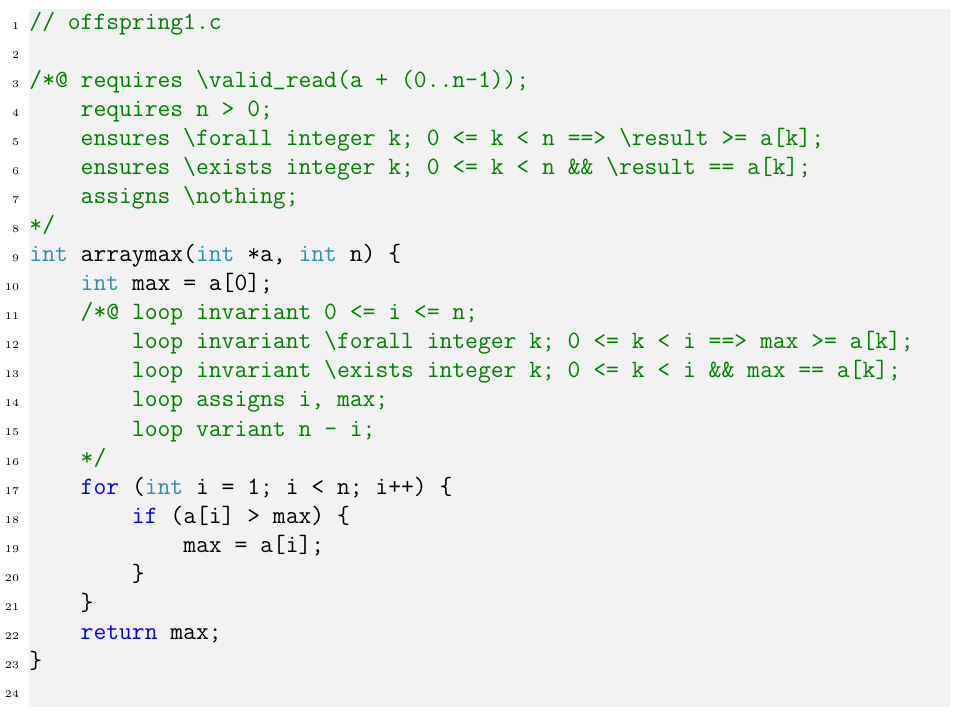}
    \end{minipage}
    \hfill
    \begin{minipage}[t]{0.49\textwidth}
        \centering
        \includegraphics[width=6.2cm,height=4.4cm]{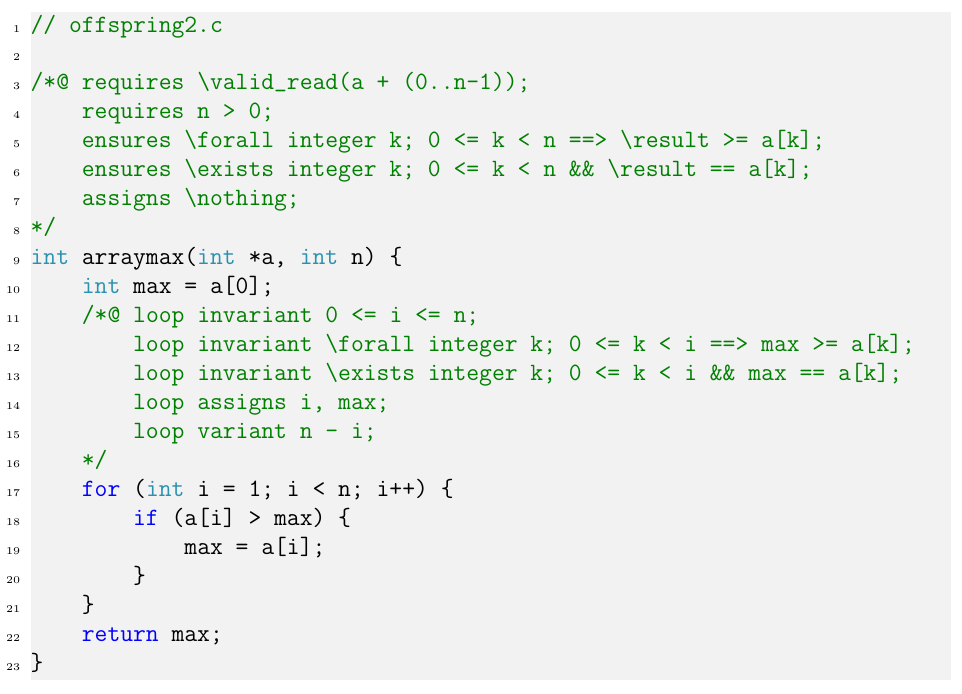}
    \end{minipage}
    
    \caption{A case of collaborative crossover.
    The top left and right subfigures are the codes in parent $\parent_1$ and $\parent_2$, respectively.
    The bottom left and right subfigures are the codes in their offspring $\offspring_1$ and $\offspring_2$, respectively.}
    \label{fig:resutl-case-collaborative-crossover}
\end{figure}

We provide two representative cases to demonstrate the utility of collaborative crossover and self-reflective mutation. 
Such cases are ubiquitous during the evolutionary process of \Solver.

The instance is sourced from \FMBENCHDF, with its natural language requirement described as follows:
`The function \code{arraymax} is designed to find the maximum value in an array of integers.
The function iterates through the array, updating the \code{max} variable whenever it finds a larger value, and finally returns the maximum value found in the array.'
As shown in Fig.~\ref{fig:resutl-case-collaborative-crossover}, the code of $\parent_1$ is successfully verified by Frama-C but fails to pass the verification of the WP plugin, while the code of $\parent_2$ fails to pass the verification of both Frama-C and the WP plugin. 

On the one hand, $\parent_1$ learns from $\parent_2$ the missing precondition \code{requires n > 0;} and loop variant clause \code{loop variant n - i;}. Additionally, $\parent_1$ recognizes and corrects the inaccurate interval bounds in the loop invariant, \ie, \code{loop invariant 0 <= i <= n}. 
This process generates the verifiable code of $\offspring_1$.
The partial thinking of qwen3-AWQ, shown as follows, demonstrates the aforementioned learning process.

\begin{llmthought}[Thinking of Refining $\parent_1$]
    Parent2's code has a requires clause that includes \code{n > 0}, which might be a necessary condition. Also, parent2 uses a \code{loop variant n - i;}, which helps in proving termination. The main issues with parent1 are: 1. Missing the \code{n > 0} condition in the requires clause to prevent division by zero or invalid array access. 2. The loop variant is missing in parent1, which could cause the WP plugin to have trouble proving termination. So to improve parent1's code: - Add a requires clause that \code{n > 0}, ensuring the array has at least one element. - Add a loop variant to help the prover with termination (like \code{n - i}).
    When the loop is done, \code{i} is \code{n}. Therefore, the loop invariant should be \code{0 <= i <= n}. Because during the loop, before each iteration, \code{i} starts at \code{1}, and after the loop, \code{i} is \code{n}. So the loop invariant should be \code{0 <= i <= n}. Therefore, adjusting the loop invariant to include \code{i <= n} would make it more accurate. The original parent1's loop invariant says \code{i < n}, which is true during the loop iterations but not after.
\end{llmthought}

On the other hand, $\parent_2$ successfully incorporates the strengths of $\parent_1$, replacing the invalid forms \verb|valid_read_only| and \verb|-->| with the correct predicate \verb|\valid_read| and implication operator \verb|==>|, respectively. 
This correction results in the verifiable code of $\offspring_2$.
The partial thinking of qwen3-AWQ, shown as follows, shows the aforementioned process.

\begin{llmthought}[Thinking of Refining $\parent_2$]
    Okay, let's start by comparing parent1 and parent2's code to see what can be improved. First, looking at parent1's full code. In the same line, parent1's code uses \code{==>} while parent2’s uses \code{--->}. In ACSL, the logical implication is written as \code{==>}, not \code{--->}. That's likely causing the parsing error. So parent2's code has invalid symbols in the ACSL annotations. So the first step is to fix the ACSL syntax in parent2's code. Replace all \code{--->} with \code{==>}.
    Now, parent1's code has a different requires clause: \code{valid\_read\_only(a + (0..n-1))}. But there's a problem here because \code{valid\_read\_only} isn't a standard ACSL predicate. That's why Frama-C gave an error, and the \code{base\_pass} is 0. So parent1's code is invalid due to a typo or incorrect predicate.
\end{llmthought}

This case demonstrates how the collaborative crossover effectively relieves the hallucination problem, wherein they tend to generate plausible but syntactically or semantically incorrect specifications. 
The mechanism cross-validates and combines complementary valid fragments from different candidate codes.

\begin{figure}[t]
    \centering
    \begin{minipage}[t]{0.49\textwidth}
        \centering
        \includegraphics[width=6.2cm,height=4.2cm]{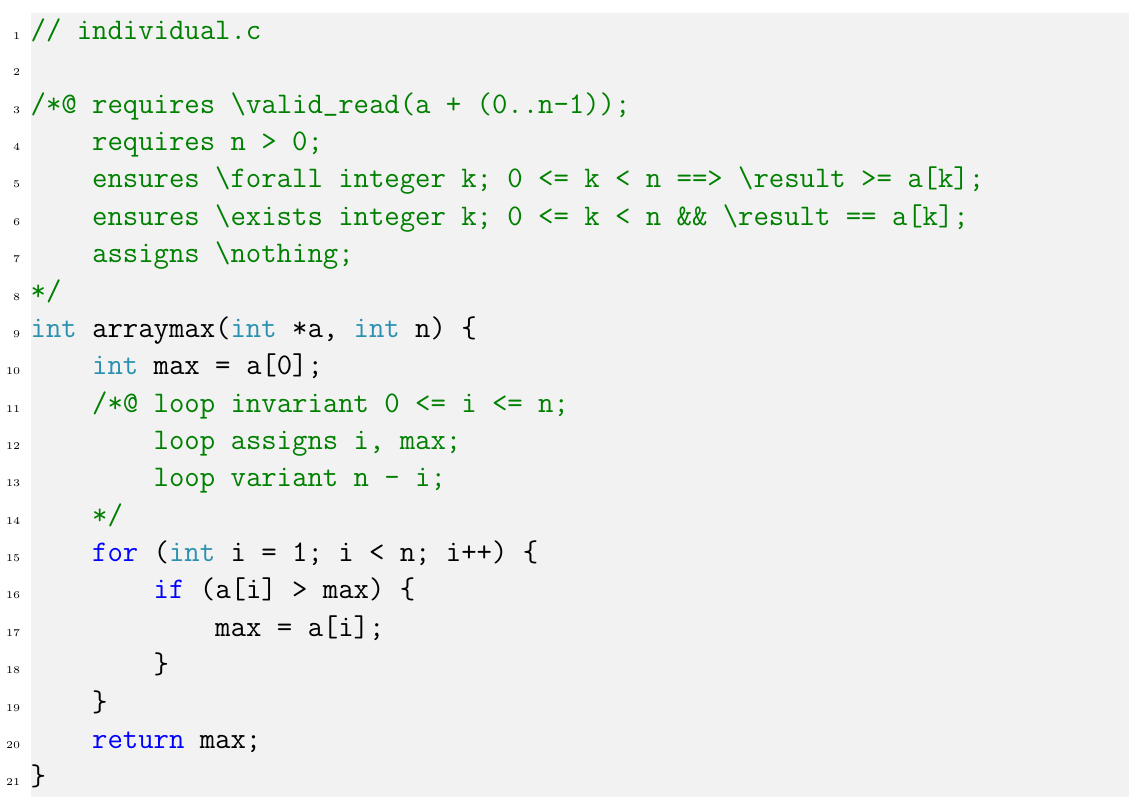}
    \end{minipage}
    \hfill
    \begin{minipage}[t]{0.49\textwidth}
        \centering
        \includegraphics[width=6.2cm,height=4.2cm]{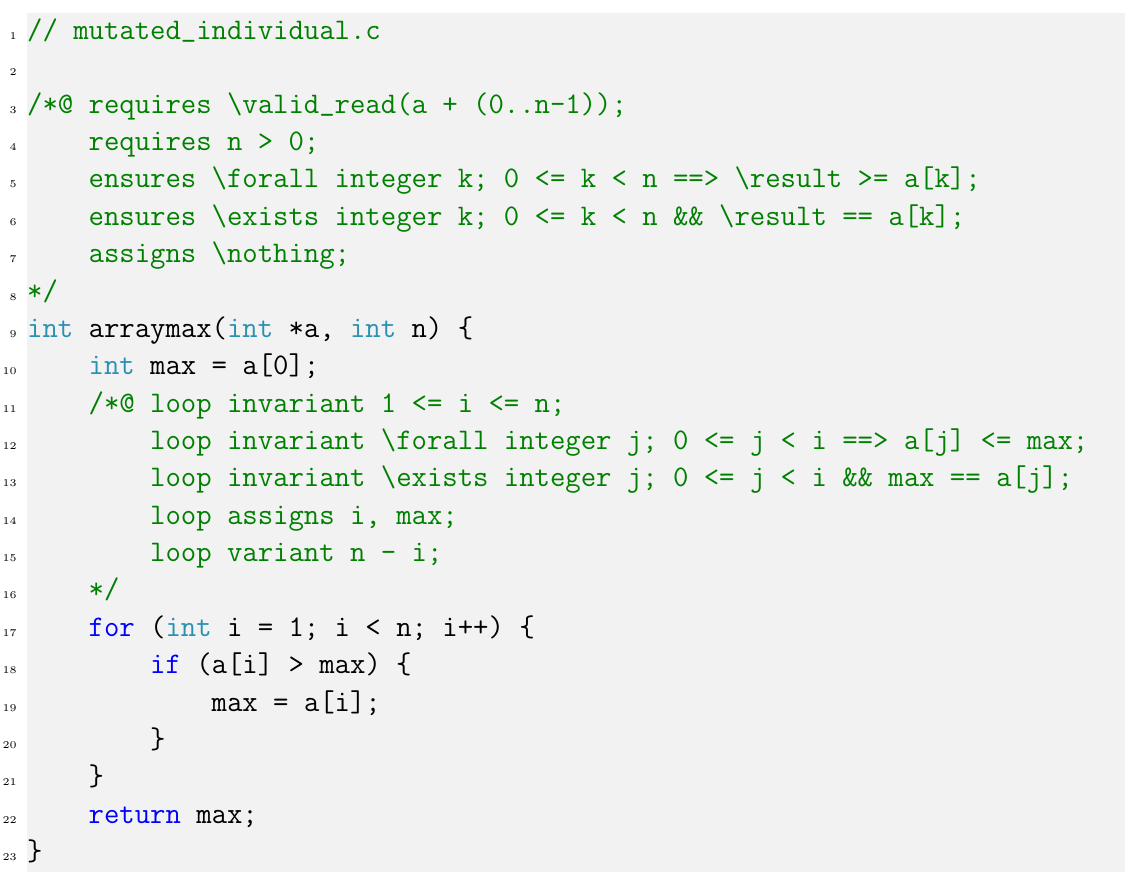}
    \end{minipage}

    \caption{An example of self-reflective mutation.
    The left and right subfigures are the codes in individual $\offspring$ and the mutated individual $\offspring'$, respectively.}
    \label{fig:resutl-case-self-reflective-mutation}
\end{figure}

As shown in Fig.~\ref{fig:resutl-case-self-reflective-mutation}, due to the missing loop invariant clauses, the code in individual $\offspring$ fails verification of the WP plugin. 
Given that its verification feedback indicated the error \code{`typed\_arraymax\_ensures'}, it 
motivates LLMs to introduce additional loop invariant clauses to support the proof of the postcondition.
After mutation, the loop invariant clauses in the mutated individual $\offspring'$ are correctly inserted, enabling it to pass verification by both Frama-C and the WP plugin.
The partial thinking of qwen3-AWQ, shown as follows, confirm the aforementioned process.

\begin{llmthought}[Thinking of Mutating $\offspring$]
    Okay, let me try to figure out how to improve the given code. The problem is in the ensures clauses. The first ensures says that the result is \code{>=} all elements. The second ensures states that the result is equal to some element in the array. The WP plugin's output shows that there are timeouts in proving these ensures clauses, which suggests that the loop invariant might not be sufficient to establish the postconditions. 
    So, the loop invariant needs to be updated. The original loop invariant doesn't state that \code{max} is the maximum of \code{a[0..i-1]}. Wait, let me think: when \code{i} starts at \code{1}, the initial \code{max} is \code{a[0]}. Then, for \code{i=1}, the loop checks \code{a[1]} against \code{max}. So after each iteration, the \code{max} should be the maximum of the first \code{i} elements. So the loop invariant should include: \code{loop invariant \textbackslash forall integer j; 0 <= j < i ==> a[j] <= max;}. Additionally, the loop invariant should also state that \code{max} is one of the elements in the array up to \code{i}. So, the second part of the postcondition requires that the result exists in the array. So the loop invariant should also have that \code{max} is one of \code{a[0..i]}. So, adding: \code{loop invariant \textbackslash exists integer j; 0 <= j < i \&\& max == a[j];}.
\end{llmthought}

This case demonstrates the effectiveness of self-reflective mutation in enabling LLM-driven code correction. 
Specifically, the model is able to interpret verifier feedback and bridge the gap between verification failure and correctness without human intervention.

\section{Conclusion and Future Works}\label{sec:conclusion}

We have investigated the autoformalizing natural language requirements into verifiable C code, a process essential for democratizing formal methods but hindered by the semantic gap between human intuition and rigorous machine logic. 
While LLMs offer promising code generation capabilities, existing approaches struggle with hallucinations, error propagation during self-reflection, and an unrealistic reliance on overly detailed requirement descriptions. 
To bridge these gaps, we have presented \Solver, a novel framework that integrates LLMs with evolutionary search. 
By modeling code synthesis as a code evolutionary process, \Solver employs the diverse individual initialization to simulate varied thinking patterns, the collaborative crossover, and the self-reflective mutation to mitigate syntactic and semantic errors as well as uncover implicit knowledge often omitted in natural language requirements. 
Extensive evaluations across varying LLM backbones and $68$ instances have demonstrated that \Solver significantly outperforms state-of-the-art approaches. 
Notably, \Solver exhibits a clear advantage on developer-friendly datasets lacking explicit requirements, proving its practical value in automating the synthesis from concise functional descriptions.

In future work, we will extend our approach to other formal languages, \eg, Dafny, and investigate more sophisticated evolutionary strategies to further enhance the synthesis performance, especially on the developer-friendly datasets.

%
%
%
\bibliographystyle{splncs04}
\bibliography{autoformalization}

@inproceedings{Cao0LMLH000QCT25,
  author       = {Jialun Cao and
                  Yaojie Lu and
                  Meiziniu Li and
                  Haoyang Ma and
                  Haokun Li and
                  Mengda He and
                  Cheng Wen and
                  Le Sun and
                  Hongyu Zhang and
                  Shengchao Qin and
                  Shing{-}Chi Cheung and
                  Cong Tian},
  title        = {From Informal to Formal - Incorporating and Evaluating LLMs on Natural
                  Language Requirements to Verifiable Formal Proofs},
  booktitle    = {{ACL} {(1)}},
  pages        = {26984--27003},
  year         = {2025}
}

@inproceedings{MurrayMBGBSLGK13,
  author       = {Toby C. Murray and
                  Daniel Matichuk and
                  Matthew Brassil and
                  Peter Gammie and
                  Timothy Bourke and
                  Sean Seefried and
                  Corey Lewis and
                  Xin Gao and
                  Gerwin Klein},
  title        = {seL4: From General Purpose to a Proof of Information Flow Enforcement},
  booktitle    = {{IEEE} Symposium on Security and Privacy},
  pages        = {415--429},
  year         = {2013}
}

@article{Leroy09,
  author       = {Xavier Leroy},
  title        = {Formal verification of a realistic compiler},
  journal      = {Commun. {ACM}},
  volume       = {52},
  number       = {7},
  pages        = {107--115},
  year         = {2009}
}

@inproceedings{FariaA23,
  author       = {Jo{\~{a}}o Pascoal Faria and
                  Rui Abreu},
  title        = {Case Studies of Development of Verified Programs with Dafny for Accessibility
                  Assessment},
  booktitle    = {{FSEN}},
  volume       = {14155},
  pages        = {25--39},
  year         = {2023}
}

@inproceedings{NobleSGS22,
  author       = {James Noble and
                  David Streader and
                  Isaac Oscar Gariano and
                  Miniruwani Samarakoon},
  title        = {More Programming Than Programming: Teaching Formal Methods in a Software
                  Engineering Programme},
  booktitle    = {{NFM}},
  volume       = {13260},
  pages        = {431--450},
  year         = {2022}
}

@inproceedings{Knight02,
  author       = {John C. Knight},
  title        = {Safety critical systems: challenges and directions},
  booktitle    = {{ICSE}},
  pages        = {547--550},
  year         = {2002}
}

@article{KleinAEMSKH14,
  author       = {Gerwin Klein and
                  June Andronick and
                  Kevin Elphinstone and
                  Toby C. Murray and
                  Thomas Sewell and
                  Rafal Kolanski and
                  Gernot Heiser},
  title        = {Comprehensive formal verification of an {OS} microkernel},
  journal      = {{ACM} Trans. Comput. Syst.},
  volume       = {32},
  number       = {1},
  pages        = {2:1--2:70},
  year         = {2014}
}

@inproceedings{LiPP24,
  author       = {Yixuan Li and
                  Julian Parsert and
                  Elizabeth Polgreen},
  title        = {Guiding Enumerative Program Synthesis with Large Language Models},
  booktitle    = {{CAV} {(2)}},
  series       = {Lecture Notes in Computer Science},
  volume       = {14682},
  pages        = {280--301},
  year         = {2024}
}

@inproceedings{WuC0W0M24,
  author       = {Guangyuan Wu and
                  Weining Cao and
                  Yuan Yao and
                  Hengfeng Wei and
                  Taolue Chen and
                  Xiaoxing Ma},
  title        = {{LLM} Meets Bounded Model Checking: Neuro-symbolic Loop Invariant
                  Inference},
  booktitle    = {ASE},
  pages        = {406--417},
  year         = {2024}
}

@article{MisuLM024,
  author       = {Md Rakib Hossain Misu and
                  Cristina V. Lopes and
                  Iris Ma and
                  James Noble},
  title        = {Towards AI-Assisted Synthesis of Verified Dafny Methods},
  journal      = {Proc. {ACM} Softw. Eng.},
  volume       = {1},
  number       = {{FSE}},
  pages        = {812--835},
  year         = {2024}
}

@inproceedings{BrownMRSKDNSSAA20,
  author       = {Tom B. Brown and
                  Benjamin Mann and
                  Nick Ryder and
                  Melanie Subbiah and
                  Jared Kaplan and
                  Prafulla Dhariwal and
                  Arvind Neelakantan and
                  Pranav Shyam and
                  Girish Sastry and
                  Amanda Askell and
                  Sandhini Agarwal and
                  Ariel Herbert{-}Voss and
                  Gretchen Krueger and
                  Tom Henighan and
                  Rewon Child and
                  Aditya Ramesh and
                  Daniel M. Ziegler and
                  Jeffrey Wu and
                  Clemens Winter and
                  Christopher Hesse and
                  Mark Chen and
                  Eric Sigler and
                  Mateusz Litwin and
                  Scott Gray and
                  Benjamin Chess and
                  Jack Clark and
                  Christopher Berner and
                  Sam McCandlish and
                  Alec Radford and
                  Ilya Sutskever and
                  Dario Amodei},
  title        = {Language Models are Few-Shot Learners},
  booktitle    = {NeurIPS},
  year         = {2020}
}

@article{ChangWWWYZCYWWYZCYYX24,
  author       = {Yupeng Chang and
                  Xu Wang and
                  Jindong Wang and
                  Yuan Wu and
                  Linyi Yang and
                  Kaijie Zhu and
                  Hao Chen and
                  Xiaoyuan Yi and
                  Cunxiang Wang and
                  Yidong Wang and
                  Wei Ye and
                  Yue Zhang and
                  Yi Chang and
                  Philip S. Yu and
                  Qiang Yang and
                  Xing Xie},
  title        = {A Survey on Evaluation of Large Language Models},
  journal      = {{ACM} Trans. Intell. Syst. Technol.},
  volume       = {15},
  number       = {3},
  pages        = {39:1--39:45},
  year         = {2024}
}

@inproceedings{ZengMYZ24,
  author       = {Jiali Zeng and
                  Fandong Meng and
                  Yongjing Yin and
                  Jie Zhou},
  title        = {Teaching Large Language Models to Translate with Comparison},
  booktitle    = {{AAAI}},
  pages        = {19488--19496},
  year         = {2024}
}

@article{abs-2107-03374,
  author       = {Mark Chen and
                  Jerry Tworek and
                  Heewoo Jun and
                  Qiming Yuan and
                  Henrique Pond{\'{e}} de Oliveira Pinto and
                  Jared Kaplan and
                  Harri Edwards and
                  Yuri Burda and
                  Nicholas Joseph and
                  Greg Brockman and
                  Alex Ray and
                  Raul Puri and
                  Gretchen Krueger and
                  Michael Petrov and
                  Heidy Khlaaf and
                  Girish Sastry and
                  Pamela Mishkin and
                  Brooke Chan and
                  Scott Gray and
                  Nick Ryder and
                  Mikhail Pavlov and
                  Alethea Power and
                  Lukasz Kaiser and
                  Mohammad Bavarian and
                  Clemens Winter and
                  Philippe Tillet and
                  Felipe Petroski Such and
                  Dave Cummings and
                  Matthias Plappert and
                  Fotios Chantzis and
                  Elizabeth Barnes and
                  Ariel Herbert{-}Voss and
                  William Hebgen Guss and
                  Alex Nichol and
                  Alex Paino and
                  Nikolas Tezak and
                  Jie Tang and
                  Igor Babuschkin and
                  Suchir Balaji and
                  Shantanu Jain and
                  William Saunders and
                  Christopher Hesse and
                  Andrew N. Carr and
                  Jan Leike and
                  Joshua Achiam and
                  Vedant Misra and
                  Evan Morikawa and
                  Alec Radford and
                  Matthew Knight and
                  Miles Brundage and
                  Mira Murati and
                  Katie Mayer and
                  Peter Welinder and
                  Bob McGrew and
                  Dario Amodei and
                  Sam McCandlish and
                  Ilya Sutskever and
                  Wojciech Zaremba},
  title        = {Evaluating Large Language Models Trained on Code},
  journal      = {CoRR},
  volume       = {abs/2107.03374},
  year         = {2021}
}

@inproceedings{LuGRHSBCDJTLZSZ21,
  author       = {Shuai Lu and
                  Daya Guo and
                  Shuo Ren and
                  Junjie Huang and
                  Alexey Svyatkovskiy and
                  Ambrosio Blanco and
                  Colin B. Clement and
                  Dawn Drain and
                  Daxin Jiang and
                  Duyu Tang and
                  Ge Li and
                  Lidong Zhou and
                  Linjun Shou and
                  Long Zhou and
                  Michele Tufano and
                  Ming Gong and
                  Ming Zhou and
                  Nan Duan and
                  Neel Sundaresan and
                  Shao Kun Deng and
                  Shengyu Fu and
                  Shujie Liu},
  title        = {CodeXGLUE: {A} Machine Learning Benchmark Dataset for Code Understanding
                  and Generation},
  booktitle    = {NeurIPS Datasets and Benchmarks},
  year         = {2021}
}

@inproceedings{WuJLRSJS22,
  author       = {Yuhuai Wu and
                  Albert Qiaochu Jiang and
                  Wenda Li and
                  Markus N. Rabe and
                  Charles Staats and
                  Mateja Jamnik and
                  Christian Szegedy},
  title        = {Autoformalization with Large Language Models},
  booktitle    = {NeurIPS},
  year         = {2022}
}

@inproceedings{SevenhuijsenEN25,
  author       = {Merlijn Sevenhuijsen and
                  Khashayar Etemadi and
                  Mattias Nyberg},
  title        = {VeCoGen: Automating Generation of Formally Verified {C} Code With
                  Large Language Models},
  booktitle    = {FormaliSE},
  pages        = {101--112},
  year         = {2025}
}

@inproceedings{CoslerHMST23,
  author       = {Matthias Cosler and
                  Christopher Hahn and
                  Daniel Mendoza and
                  Frederik Schmitt and
                  Caroline Trippel},
  title        = {nl2spec: Interactively Translating Unstructured Natural Language to
                  Temporal Logics with Large Language Models},
  booktitle    = {CAV},
  pages        = {383--396},
  year         = {2023}
}

@inproceedings{MolinadA23,
  author       = {Facundo Molina and
                  Marcelo d'Amorim and
                  Nazareno Aguirre},
  title        = {SpecFuzzer: {A} Tool for Inferring Class Specifications via Grammar-Based
                  Fuzzing},
  booktitle    = {{ASE}},
  pages        = {2094--2097},
  year         = {2023}
}

@inproceedings{LinCHWLLSL24,
  author       = {Xiaohan Lin and
                  Qingxing Cao and
                  Yinya Huang and
                  Haiming Wang and
                  Jianqiao Lu and
                  Zhengying Liu and
                  Linqi Song and
                  Xiaodan Liang},
  title        = {{FVEL:} Interactive Formal Verification Environment with Large Language
                  Models via Theorem Proving},
  booktitle    = {NeurIPS},
  year         = {2024}
}

@inproceedings{RamanathanGJ07,
  author       = {Murali Krishna Ramanathan and
                  Ananth Grama and
                  Suresh Jagannathan},
  title        = {Static specification inference using predicate mining},
  booktitle    = {{PLDI}},
  pages        = {123--134},
  year         = {2007}
}

@inproceedings{BeckmanN11,
  author       = {Nels E. Beckman and
                  Aditya V. Nori},
  title        = {Probabilistic, modular and scalable inference of typestate specifications},
  booktitle    = {{PLDI}},
  pages        = {211--221},
  year         = {2011}
}

@article{ShohamYFP08,
  author       = {Sharon Shoham and
                  Eran Yahav and
                  Stephen J. Fink and
                  Marco Pistoia},
  title        = {Static Specification Mining Using Automata-Based Abstractions},
  journal      = {{IEEE} Trans. Software Eng.},
  volume       = {34},
  number       = {5},
  pages        = {651--666},
  year         = {2008}
}

@inproceedings{WenCSXQHLCT24,
  author       = {Cheng Wen and
                  Jialun Cao and
                  Jie Su and
                  Zhiwu Xu and
                  Shengchao Qin and
                  Mengda He and
                  Haokun Li and
                  Shing{-}Chi Cheung and
                  Cong Tian},
  title        = {Enchanting Program Specification Synthesis by Large Language Models
                  Using Static Analysis and Program Verification},
  booktitle    = {CAV},
  volume       = {14682},
  pages        = {302--328},
  year         = {2024}
}

@inproceedings{FirstRRB23,
  author       = {Emily First and
                  Markus N. Rabe and
                  Talia Ringer and
                  Yuriy Brun},
  title        = {Baldur: Whole-Proof Generation and Repair with Large Language Models},
  booktitle    = {{FSE}},
  pages        = {1229--1241},
  year         = {2023}
}

@inproceedings{YangSGCSYGPA23,
  author       = {Kaiyu Yang and
                  Aidan M. Swope and
                  Alex Gu and
                  Rahul Chalamala and
                  Peiyang Song and
                  Shixing Yu and
                  Saad Godil and
                  Ryan J. Prenger and
                  Animashree Anandkumar},
  title        = {LeanDojo: Theorem Proving with Retrieval-Augmented Language Models},
  booktitle    = {NeurIPS},
  year         = {2023}
}

@inproceedings{LuSSHZJLT25,
  author       = {Xu Lu and
                  Weisong Sun and
                  Yiran Zhang and
                  Ming Hu and
                  Cong Tian and
                  Zhi Jin and
                  Yang Liu},
  title        = {Requirements Development and Formalization for Reliable Code Generation:
                  {A} Multi-Agent Vision},
  booktitle    = {ASE},
  year         = {2025}
}

@article{abs-2410-14835,
  author       = {Prasita Mukherjee and
                  Benjamin Delaware},
  title        = {Towards Automated Verification of LLM-Synthesized {C} Programs},
  journal      = {CoRR},
  volume       = {abs/2410.14835},
  year         = {2024}
}

@inproceedings{ZhaiSPZLFM0020,
  author       = {Juan Zhai and
                  Yu Shi and
                  Minxue Pan and
                  Guian Zhou and
                  Yongxiang Liu and
                  Chunrong Fang and
                  Shiqing Ma and
                  Lin Tan and
                  Xiangyu Zhang},
  title        = {{C2S:} translating natural language comments to formal program specifications},
  booktitle    = {{FSE}},
  pages        = {25--37},

  year         = {2020}
}

@article{HembergMO24,
  author       = {Erik Hemberg and
                  Stephen Moskal and
                  Una{-}May O'Reilly},
  title        = {Evolving code with a large language model},
  journal      = {Genet. Program. Evolvable Mach.},
  volume       = {25},
  number       = {2},
  pages        = {21},
  year         = {2024}
}

@inproceedings{DatDB25,
  author       = {Pham Vu Tuan Dat and
                  Long Doan and
                  Huynh Thi Thanh Binh},
  title        = {HSEvo: Elevating Automatic Heuristic Design with Diversity-Driven
                  Harmony Search and Genetic Algorithm Using LLMs},
  booktitle    = {{AAAI}},
  pages        = {26931--26938},
  year         = {2025}
}

@article{abs-2406-07496,
  author       = {Mert Y{\"{u}}ksekg{\"{o}}n{\"{u}}l and
                  Federico Bianchi and
                  Joseph Boen and
                  Sheng Liu and
                  Zhi Huang and
                  Carlos Guestrin and
                  James Zou},
  title        = {TextGrad: Automatic "Differentiation" via Text},
  journal      = {CoRR},
  volume       = {abs/2406.07496},
  year         = {2024}
}

@inproceedings{Guo0GLS0L0Y24,
  author       = {Qingyan Guo and
                  Rui Wang and
                  Junliang Guo and
                  Bei Li and
                  Kaitao Song and
                  Xu Tan and
                  Guoqing Liu and
                  Jiang Bian and
                  Yujiu Yang},
  title        = {Connecting Large Language Models with Evolutionary Algorithms Yields
                  Powerful Prompt Optimizers},
  booktitle    = {{ICLR}},
  year         = {2024}
}

@article{NejjarZSW25,
  author       = {Mohamed Nejjar and
                  Luca Zacharias and
                  Fabian Stiehle and
                  Ingo Weber},
  title        = {LLMs for science: Usage for code generation and data analysis},
  journal      = {J. Softw. Evol. Process.},
  volume       = {37},
  number       = {1},
  year         = {2025}
}

@article{RomeraParedesBNBKDREWFKF24,
  author       = {Bernardino Romera{-}Paredes and
                  Mohammadamin Barekatain and
                  Alexander Novikov and
                  Matej Balog and
                  M. Pawan Kumar and
                  Emilien Dupont and
                  Francisco J. R. Ruiz and
                  Jordan S. Ellenberg and
                  Pengming Wang and
                  Omar Fawzi and
                  Pushmeet Kohli and
                  Alhussein Fawzi},
  title        = {Mathematical discoveries from program search with large language models},
  journal      = {Nat.},
  volume       = {625},
  number       = {7995},
  pages        = {468--475},
  year         = {2024}
}

@inproceedings{0044TY0LWL024,
  author       = {Fei Liu and
                  Xialiang Tong and
                  Mingxuan Yuan and
                  Xi Lin and
                  Fu Luo and
                  Zhenkun Wang and
                  Zhichao Lu and
                  Qingfu Zhang},
  title        = {Evolution of Heuristics: Towards Efficient Automatic Algorithm Design
                  Using Large Language Model},
  booktitle    = {{ICML}},
  year         = {2024}
}

@inproceedings{Ye0CBHKPS24,
  author       = {Haoran Ye and
                  Jiarui Wang and
                  Zhiguang Cao and
                  Federico Berto and
                  Chuanbo Hua and
                  Haeyeon Kim and
                  Jinkyoo Park and
                  Guojie Song},
  title        = {ReEvo: Large Language Models as Hyper-Heuristics with Reflective Evolution},
  booktitle    = {NeurIPS},
  year         = {2024}
}

@inproceedings{LeQC15,
  author       = {Ton Chanh Le and
                  Shengchao Qin and
                  Wei{-}Ngan Chin},
  title        = {Termination and non-termination specification inference},
  booktitle    = {{PLDI}},
  pages        = {489--498},
  year         = {2015}
}

@inproceedings{Yu0023,
  author       = {Shiwen Yu and
                  Ting Wang and
                  Ji Wang},
  title        = {Loop Invariant Inference through {SMT} Solving Enhanced Reinforcement
                  Learning},
  booktitle    = {{ISSTA}},
  pages        = {175--187},
  year         = {2023}
}

@inproceedings{LinZCSXLS21,
  author       = {Yingwen Lin and
                  Yao Zhang and
                  Sen Chen and
                  Fu Song and
                  Xiaofei Xie and
                  Xiaohong Li and
                  Lintan Sun},
  title        = {Inferring Loop Invariants for Multi-Path Loops},
  booktitle    = {{TASE}},
  pages        = {63--70},
  year         = {2021}
}

@inproceedings{DilligDLM13,
  author       = {Isil Dillig and
                  Thomas Dillig and
                  Boyang Li and
                  Kenneth L. McMillan},
  title        = {Inductive invariant generation via abductive inference},
  booktitle    = {{OOPSLA}},
  pages        = {443--456},
  year         = {2013}
}

@inproceedings{RyanWYGJ20,
  author       = {Gabriel Ryan and
                  Justin Wong and
                  Jianan Yao and
                  Ronghui Gu and
                  Suman Jana},
  title        = {{CLN2INV:} Learning Loop Invariants with Continuous Logic Networks},
  booktitle    = {{ICLR}},
  year         = {2020}
}

@inproceedings{SiDRNS18,
  author       = {Xujie Si and
                  Hanjun Dai and
                  Mukund Raghothaman and
                  Mayur Naik and
                  Le Song},
  title        = {Learning Loop Invariants for Program Verification},
  booktitle    = {NeurIPS},
  pages        = {7762--7773},
  year         = {2018}
}

@inproceedings{YaoRWJG20,
  author       = {Jianan Yao and
                  Gabriel Ryan and
                  Justin Wong and
                  Suman Jana and
                  Ronghui Gu},
  title        = {Learning nonlinear loop invariants with gated continuous logic networks},
  booktitle    = {{PLDI}},
  pages        = {106--120},
  year         = {2020}
}

@inproceedings{LinSXLSH17,
  author       = {Shang{-}Wei Lin and
                  Jun Sun and
                  Hao Xiao and
                  Yang Liu and
                  David San{\'{a}}n and
                  Henri Hansen},
  title        = {FiB: squeezing loop invariants by interpolation between Forward/Backward
                  predicate transformers},
  booktitle    = {{ASE}},
  pages        = {793--803},

  year         = {2017}
}

@inproceedings{0001BN24,
  author       = {Haoze Wu and
                  Clark W. Barrett and
                  Nina Narodytska},
  title        = {Lemur: Integrating Large Language Models in Automated Program Verification},
  booktitle    = {{ICLR}},
  year         = {2024}
}

@inproceedings{PeiBSSY23,
  author       = {Kexin Pei and
                  David Bieber and
                  Kensen Shi and
                  Charles Sutton and
                  Pengcheng Yin},
  title        = {Can Large Language Models Reason about Program Invariants?},
  booktitle    = {{ICML}},
  volume       = {202},
  pages        = {27496--27520},
  year         = {2023}
}

@article{CaoWXYWCM25,
  author       = {Weining Cao and
                  Guangyuan Wu and
                  Tangzhi Xu and
                  Yuan Yao and
                  Hengfeng Wei and
                  Taolue Chen and
                  Xiaoxing Ma},
  title        = {Clause2Inv: {A} Generate-Combine-Check Framework for Loop Invariant
                  Inference},
  journal      = {Proc. {ACM} Softw. Eng.},
  volume       = {2},
  number       = {{ISSTA}},
  pages        = {1009--1030},
  year         = {2025}
}

@article{YuLWW24,
  author       = {Shiwen Yu and
                  Zengyu Liu and
                  Ting Wang and
                  Ji Wang},
  title        = {Neural Solving Uninterpreted Predicates with Abstract Gradient Descent},
  journal      = {{ACM} Trans. Softw. Eng. Methodol.},
  volume       = {33},
  number       = {8},
  pages        = {215:1--215:47},
  year         = {2024}
}

@inproceedings{PirzadaRBC24,
  author       = {Muhammad A. A. Pirzada and
                  Giles Reger and
                  Ahmed Bhayat and
                  Lucas C. Cordeiro},
  title        = {LLM-Generated Invariants for Bounded Model Checking Without Loop Unrolling},
  booktitle    = {{ASE}},
  pages        = {1395--1407},
  year         = {2024}
}

@article{KarbyshevBIRS17,
  author       = {Aleksandr Karbyshev and
                  Nikolaj S. Bj{\o}rner and
                  Shachar Itzhaky and
                  Noam Rinetzky and
                  Sharon Shoham},
  title        = {Property-Directed Inference of Universal Invariants or Proving Their
                  Absence},
  journal      = {J. {ACM}},
  volume       = {64},
  number       = {1},
  pages        = {7:1--7:33},
  year         = {2017}
}

@article{abs-2509-09917,
  author       = {Zehan Chen and
                  Long Zhang and
                  Zhiwei Zhang and
                  JingJing Zhang and
                  Ruoyu Zhou and
                  Yulong Shen and
                  Jianfeng Ma and
                  Lin Yang},
  title        = {SLD-Spec: Enhancement LLM-assisted Specification Generation for Complex
                  Loop Functions via Program Slicing and Logical Deletion},
  journal      = {CoRR},
  volume       = {abs/2509.09917},
  year         = {2025}
}

@inproceedings{CousotCFL13,
  author       = {Patrick Cousot and
                  Radhia Cousot and
                  Manuel F{\"{a}}hndrich and
                  Francesco Logozzo},
  title        = {Automatic Inference of Necessary Preconditions},
  booktitle    = {{VMCAI}},
  volume       = {7737},
  pages        = {128--148},
  year         = {2013}
}

@inproceedings{PadhiSM16,
  author       = {Saswat Padhi and
                  Rahul Sharma and
                  Todd D. Millstein},
  title        = {Data-driven precondition inference with learned features},
  booktitle    = {{PLDI}},
  pages        = {42--56},
  year         = {2016}
}

@inproceedings{PopeeaC06,
  author       = {Corneliu Popeea and
                  Wei{-}Ngan Chin},
  title        = {Inferring Disjunctive Postconditions},
  booktitle    = {{ASIAN}},
  volume       = {4435},
  pages        = {331--345},
  year         = {2006}
}

@inproceedings{SuAD18,
  author       = {Jingyi Su and
                  Mohd Arafat and
                  Robert Dyer},
  title        = {Using consensus to automatically infer post-conditions},
  booktitle    = {{ICSE} (Companion Volume)},
  pages        = {202--203},
  year         = {2018}
}

@inproceedings{SingletonLRC18,
  author       = {John L. Singleton and
                  Gary T. Leavens and
                  Hridesh Rajan and
                  David R. Cok},
  title        = {An algorithm and tool to infer practical postconditions},
  booktitle    = {{ICSE} (Companion Volume)},
  pages        = {313--314},
  year         = {2018}
}

@article{KirchnerKPSY15,
  author       = {Florent Kirchner and
                  Nikolai Kosmatov and
                  Virgile Prevosto and
                  Julien Signoles and
                  Boris Yakobowski},
  title        = {Frama-C: {A} software analysis perspective},
  journal      = {Formal Aspects Comput.},
  volume       = {27},
  number       = {3},
  pages        = {573--609},
  year         = {2015}
}

@inproceedings{CuoqKKPSY12,
  author       = {Pascal Cuoq and
                  Florent Kirchner and
                  Nikolai Kosmatov and
                  Virgile Prevosto and
                  Julien Signoles and
                  Boris Yakobowski},
  title        = {Frama-C - {A} Software Analysis Perspective},
  booktitle    = {{SEFM}},
  volume       = {7504},
  pages        = {233--247},
  year         = {2012}
}

@inproceedings{Dordowsky15,
  author       = {Frank Dordowsky},
  title        = {An experimental Study using {ACSL} and Frama-C to formulate and verify
                  Low-Level Requirements from a {DO-178C} compliant Avionics Project},
  booktitle    = {{F-IDE}},
  volume       = {187},
  pages        = {28--41},
  year         = {2015}
}

@inproceedings{BritoP10,
  author       = {Eduardo Brito and
                  Jorge Sousa Pinto},
  title        = {Program Verification in {SPARK} and {ACSL:} {A} Comparative Case Study},
  booktitle    = {Ada-Europe},
  volume       = {6106},
  pages        = {97--110},
  year         = {2010}
}

@inproceedings{UngAGLNP24,
  author       = {Gustav Ung and
                  Jesper Amilon and
                  Dilian Gurov and
                  Christian Lidstr{\"{o}}m and
                  Mattias Nyberg and
                  Karl Palmskog},
  title        = {Post-Hoc Formal Verification of Automotive Software with Informal
                  Requirements: An Experience Report},
  booktitle    = {{RE}},
  pages        = {287--298},
  year         = {2024}
}

@article{Hoare69,
  author       = {C. A. R. Hoare},
  title        = {An Axiomatic Basis for Computer Programming},
  journal      = {Commun. {ACM}},
  volume       = {12},
  number       = {10},
  pages        = {576--580},
  year         = {1969}
}

@techreport{Filliatre2003multi,
  title={Why: a multi-language multi-prover verification tool},
  author={Filli{\^a}tre, Jean-Christophe},
  year={2003},
  institution={Research Report 1366, LRI, Universit{\'e} Paris Sud}
}

@misc{Baudin2021acsl,
  title={Acsl: Ansi/iso c specification},
  author={Baudin, Patrick and Filli{\^a}tre, Jean-Christophe and March{\'e}, Claude and Monate, Benjamin and Moy, Yannick and Prevosto, Virgile},
  howpublished = {\url{https://frama-c.com/html/acsl.html}},
  year={2021}
}

@inproceedings{MouraB08,
  author       = {Leonardo Mendon{\c{c}}a de Moura and
                  Nikolaj S. Bj{\o}rner},
  title        = {{Z3:} An Efficient {SMT} Solver},
  booktitle    = {{TACAS}},
  volume       = {4963},
  pages        = {337--340},
  year         = {2008}
}

@inproceedings{VaswaniSPUJGKP17,
  author       = {Ashish Vaswani and
                  Noam Shazeer and
                  Niki Parmar and
                  Jakob Uszkoreit and
                  Llion Jones and
                  Aidan N. Gomez and
                  Lukasz Kaiser and
                  Illia Polosukhin},
  title        = {Attention is All you Need},
  booktitle    = {{NIPS}},
  pages        = {5998--6008},
  year         = {2017}
}

@inproceedings{LiLZFCLC23,
  author       = {Yifei Li and
                  Zeqi Lin and
                  Shizhuo Zhang and
                  Qiang Fu and
                  Bei Chen and
                  Jian{-}Guang Lou and
                  Weizhu Chen},
  title        = {Making Language Models Better Reasoners with Step-Aware Verifier},
  booktitle    = {{ACL} {(1)}},
  pages        = {5315--5333},
  year         = {2023}
}

@inproceedings{BachSYWRNSKBFAD22,
  author       = {Stephen H. Bach and
                  Victor Sanh and
                  Zheng Xin Yong and
                  Albert Webson and
                  Colin Raffel and
                  Nihal V. Nayak and
                  Abheesht Sharma and
                  Taewoon Kim and
                  M. Saiful Bari and
                  Thibault F{\'{e}}vry and
                  Zaid Alyafeai and
                  Manan Dey and
                  Andrea Santilli and
                  Zhiqing Sun and
                  Srulik Ben{-}David and
                  Canwen Xu and
                  Gunjan Chhablani and
                  Han Wang and
                  Jason Alan Fries and
                  Maged Saeed AlShaibani and
                  Shanya Sharma and
                  Urmish Thakker and
                  Khalid Almubarak and
                  Xiangru Tang and
                  Dragomir R. Radev and
                  Mike Tian{-}Jian Jiang and
                  Alexander M. Rush},
  title        = {PromptSource: An Integrated Development Environment and Repository
                  for Natural Language Prompts},
  booktitle    = {{ACL} (demo)},
  pages        = {93--104},
  year         = {2022}
}

@article{abs-2209-01390,
  author       = {Hai Dang and
                  Lukas Mecke and
                  Florian Lehmann and
                  Sven Goller and
                  Daniel Buschek},
  title        = {How to Prompt? Opportunities and Challenges of Zero- and Few-Shot
                  Learning for Human-AI Interaction in Creative Applications of Generative
                  Models},
  journal      = {CoRR},
  volume       = {abs/2209.01390},
  year         = {2022}
}

@inproceedings{DeckersFKPS0P23,
  author       = {Niklas Deckers and
                  Maik Fr{\"{o}}be and
                  Johannes Kiesel and
                  Gianluca Pandolfo and
                  Christopher Schr{\"{o}}der and
                  Benno Stein and
                  Martin Potthast},
  title        = {The Infinite Index: Information Retrieval on Generative Text-To-Image
                  Models},
  booktitle    = {{CHIIR}},
  pages        = {172--186},
  year         = {2023}
}

@inproceedings{HouDWLC22,
  author       = {Yutai Hou and
                  Hongyuan Dong and
                  Xinghao Wang and
                  Bohan Li and
                  Wanxiang Che},
  title        = {MetaPrompting: Learning to Learn Better Prompts},
  booktitle    = {{COLING}},
  pages        = {3251--3262},
  year         = {2022}
}

@inproceedings{JiangOTMDTC22,
  author       = {Ellen Jiang and
                  Kristen Olson and
                  Edwin Toh and
                  Alejandra Molina and
                  Aaron Donsbach and
                  Michael Terry and
                  Carrie J. Cai},
  title        = {PromptMaker: Prompt-based Prototyping with Large Language Models},
  booktitle    = {{CHI} Extended Abstracts},
  pages        = {35:1--35:8},
  year         = {2022}
}

@inproceedings{LiuC22a,
  author       = {Vivian Liu and
                  Lydia B. Chilton},
  title        = {Design Guidelines for Prompt Engineering Text-to-Image Generative
                  Models},
  booktitle    = {{CHI}},
  pages        = {384:1--384:23},
  year         = {2022}
}

@article{Beurer-Kellner023,
  author       = {Luca Beurer{-}Kellner and
                  Marc Fischer and
                  Martin T. Vechev},
  title        = {Prompting Is Programming: {A} Query Language for Large Language Models},
  journal      = {Proc. {ACM} Program. Lang.},
  volume       = {7},
  number       = {{PLDI}},
  pages        = {1946--1969},
  year         = {2023}
}

@inproceedings{ReynoldsM21,
  author       = {Laria Reynolds and
                  Kyle McDonell},
  title        = {Prompt Programming for Large Language Models: Beyond the Few-Shot
                  Paradigm},
  booktitle    = {{CHI} Extended Abstracts},
  pages        = {314:1--314:7},
  year         = {2021}
}

@misc{Conchon2006Ergo,
  title        = {{Ergo}: A Theorem Prover for Polymorphic First-Order Logic Modulo Theories},
  author       = {Sylvain Conchon and Evelyne Contejean and Johannes Kanig},
  year         = {2006}
}

@inproceedings{BarrettCDHJKRT11,
  author       = {Clark W. Barrett and
                  Christopher L. Conway and
                  Morgan Deters and
                  Liana Hadarean and
                  Dejan Jovanovic and
                  Tim King and
                  Andrew Reynolds and
                  Cesare Tinelli},
  title        = {{CVC4}},
  booktitle    = {{CAV}},
  volume       = {6806},
  pages        = {171--177},
  year         = {2011}
}

@misc{CoqRefManual8.4,
  author       = {{Coq Development Team}},
  title        = {The {Coq} Proof Assistant Reference Manual -- Version 8.4},
  year         = {2010}
}

@inproceedings{Huet87,
  author       = {G{\'{e}}rard P. Huet},
  title        = {The Calculus of Constructions: State of the Art},
  booktitle    = {{FSTTCS}},
  volume       = {287},
  pages        = {372},
  year         = {1987}
}

@inproceedings{Moura021,
  author       = {Leonardo de Moura and
                  Sebastian Ullrich},
  title        = {The Lean 4 Theorem Prover and Programming Language},
  booktitle    = {{CADE}},
  volume       = {12699},
  pages        = {625--635},
  year         = {2021}
}

@inproceedings{Leino10,
  author       = {K. Rustan M. Leino},
  title        = {Dafny: An Automatic Program Verifier for Functional Correctness},
  booktitle    = {{LPAR} (Dakar)},
  volume       = {6355},
  pages        = {348--370},
  year         = {2010}
}

@inproceedings{YuML99,
  author       = {Yuan Yu and
                  Panagiotis Manolios and
                  Leslie Lamport},
  title        = {Model Checking TLA\({}^{\mbox{+}}\) Specifications},
  booktitle    = {{CHARME}},
  volume       = {1703},
  pages        = {54--66},
  year         = {1999}
}

@book{Lamport2002,
  author       = {Leslie Lamport},
  title        = {Specifying Systems, The {TLA+} Language and Tools for Hardware and
                  Software Engineers},
  year         = {2002}
}

@misc{Meta2024Llama3,
  author       = {{Meta AI}},
  title        = {Introducing {Meta} {Llama} 3: The most capable openly available {LLM} to date},
  year         = {2024},
  howpublished = {\url{https://ai.meta.com/blog/meta-llama-3/}}
}

@article{abs-2412-15115,
  author       = {An Yang and
                  Baosong Yang and
                  Beichen Zhang and
                  Binyuan Hui and
                  Bo Zheng and
                  Bowen Yu and
                  Chengyuan Li and
                  Dayiheng Liu and
                  Fei Huang and
                  Haoran Wei and
                  Huan Lin and
                  Jian Yang and
                  Jianhong Tu and
                  Jianwei Zhang and
                  Jianxin Yang and
                  Jiaxi Yang and
                  Jingren Zhou and
                  Junyang Lin and
                  Kai Dang and
                  Keming Lu and
                  Keqin Bao and
                  Kexin Yang and
                  Le Yu and
                  Mei Li and
                  Mingfeng Xue and
                  Pei Zhang and
                  Qin Zhu and
                  Rui Men and
                  Runji Lin and
                  Tianhao Li and
                  Tingyu Xia and
                  Xingzhang Ren and
                  Xuancheng Ren and
                  Yang Fan and
                  Yang Su and
                  Yichang Zhang and
                  Yu Wan and
                  Yuqiong Liu and
                  Zeyu Cui and
                  Zhenru Zhang and
                  Zihan Qiu},
  title        = {Qwen2.5 Technical Report},
  journal      = {CoRR},
  volume       = {abs/2412.15115},
  year         = {2024}
}

@article{abs-2409-12186,
  author       = {Binyuan Hui and
                  Jian Yang and
                  Zeyu Cui and
                  Jiaxi Yang and
                  Dayiheng Liu and
                  Lei Zhang and
                  Tianyu Liu and
                  Jiajun Zhang and
                  Bowen Yu and
                  Kai Dang and
                  An Yang and
                  Rui Men and
                  Fei Huang and
                  Xingzhang Ren and
                  Xuancheng Ren and
                  Jingren Zhou and
                  Junyang Lin},
  title        = {Qwen2.5-Coder Technical Report},
  journal      = {CoRR},
  volume       = {abs/2409.12186},
  year         = {2024}
}

@article{abs-2501-12948,
  author       = {DeepSeek{-}AI},
  title        = {DeepSeek-R1: Incentivizing Reasoning Capability in LLMs via Reinforcement
                  Learning},
  journal      = {CoRR},
  volume       = {abs/2501.12948},
  year         = {2025}
}

@inproceedings{DingCXQHL0Z23,
  author       = {Ning Ding and
                  Yulin Chen and
                  Bokai Xu and
                  Yujia Qin and
                  Shengding Hu and
                  Zhiyuan Liu and
                  Maosong Sun and
                  Bowen Zhou},
  title        = {Enhancing Chat Language Models by Scaling High-quality Instructional
                  Conversations},
  booktitle    = {{EMNLP}},
  pages        = {3029--3051},
  year         = {2023}
}

@article{abs-2411-15124,
  author       = {Nathan Lambert and
                  Jacob Morrison and
                  Valentina Pyatkin and
                  Shengyi Huang and
                  Hamish Ivison and
                  Faeze Brahman and
                  Lester James V. Miranda and
                  Alisa Liu and
                  Nouha Dziri and
                  Shane Lyu and
                  Yuling Gu and
                  Saumya Malik and
                  Victoria Graf and
                  Jena D. Hwang and
                  Jiangjiang Yang and
                  Ronan Le Bras and
                  Oyvind Tafjord and
                  Chris Wilhelm and
                  Luca Soldaini and
                  Noah A. Smith and
                  Yizhong Wang and
                  Pradeep Dasigi and
                  Hannaneh Hajishirzi},
  title        = {T{\"{U}}LU 3: Pushing Frontiers in Open Language Model Post-Training},
  journal      = {CoRR},
  volume       = {abs/2411.15124},
  year         = {2024}
}

@article{abs-2505-09388,
  author       = {An Yang and
                  Anfeng Li and
                  Baosong Yang and
                  Beichen Zhang and
                  Binyuan Hui and
                  Bo Zheng and
                  Bowen Yu and
                  Chang Gao and
                  Chengen Huang and
                  Chenxu Lv and
                  Chujie Zheng and
                  Dayiheng Liu and
                  Fan Zhou and
                  Fei Huang and
                  Feng Hu and
                  Hao Ge and
                  Haoran Wei and
                  Huan Lin and
                  Jialong Tang and
                  Jian Yang and
                  Jianhong Tu and
                  Jianwei Zhang and
                  Jian Yang and
                  Jiaxi Yang and
                  Jingren Zhou and
                  Junyang Lin and
                  Kai Dang and
                  Keqin Bao and
                  Kexin Yang and
                  Le Yu and
                  Lianghao Deng and
                  Mei Li and
                  Mingfeng Xue and
                  Mingze Li and
                  Pei Zhang and
                  Peng Wang and
                  Qin Zhu and
                  Rui Men and
                  Ruize Gao and
                  Shixuan Liu and
                  Shuang Luo and
                  Tianhao Li and
                  Tianyi Tang and
                  Wenbiao Yin and
                  Xingzhang Ren and
                  Xinyu Wang and
                  Xinyu Zhang and
                  Xuancheng Ren and
                  Yang Fan and
                  Yang Su and
                  Yichang Zhang and
                  Yinger Zhang and
                  Yu Wan and
                  Yuqiong Liu and
                  Zekun Wang and
                  Zeyu Cui and
                  Zhenru Zhang and
                  Zhipeng Zhou and
                  Zihan Qiu},
  title        = {Qwen3 Technical Report},
  journal      = {CoRR},
  volume       = {abs/2505.09388},
  year         = {2025}
}

@article{abs-2507-06261,
  author       = {Gemini Team},
  title        = {Gemini 2.5: Pushing the Frontier with Advanced Reasoning, Multimodality,
                  Long Context, and Next Generation Agentic Capabilities},
  journal      = {CoRR},
  volume       = {abs/2507.06261},
  year         = {2025}
}

@article{abs-2410-21276,
  author       = {Aaron Hurst and
                  Adam Lerer and
                  Adam P. Goucher and
                  Adam Perelman and
                  Aditya Ramesh and
                  Aidan Clark and
                  AJ Ostrow and
                  Akila Welihinda and
                  Alan Hayes and
                  Alec Radford and
                  Aleksander Madry and
                  Alex Baker{-}Whitcomb and
                  Alex Beutel and
                  Alex Borzunov and
                  Alex Carney and
                  Alex Chow and
                  Alex Kirillov and
                  Alex Nichol and
                  Alex Paino and
                  Alex Renzin and
                  Alex Tachard Passos and
                  Alexander Kirillov and
                  Alexi Christakis and
                  Alexis Conneau and
                  Ali Kamali and
                  Allan Jabri and
                  Allison Moyer and
                  Allison Tam and
                  Amadou Crookes and
                  Amin Tootoonchian and
                  Ananya Kumar and
                  Andrea Vallone and
                  Andrej Karpathy and
                  Andrew Braunstein and
                  Andrew Cann and
                  Andrew Codispoti and
                  Andrew Galu and
                  Andrew Kondrich and
                  Andrew Tulloch and
                  Andrey Mishchenko and
                  Angela Baek and
                  Angela Jiang and
                  Antoine Pelisse and
                  Antonia Woodford and
                  Anuj Gosalia and
                  Arka Dhar and
                  Ashley Pantuliano and
                  Avi Nayak and
                  Avital Oliver and
                  Barret Zoph and
                  Behrooz Ghorbani and
                  Ben Leimberger and
                  Ben Rossen and
                  Ben Sokolowsky and
                  Ben Wang and
                  Benjamin Zweig and
                  Beth Hoover and
                  Blake Samic and
                  Bob McGrew and
                  Bobby Spero and
                  Bogo Giertler and
                  Bowen Cheng and
                  Brad Lightcap and
                  Brandon Walkin and
                  Brendan Quinn and
                  Brian Guarraci and
                  Brian Hsu and
                  Bright Kellogg and
                  Brydon Eastman and
                  Camillo Lugaresi and
                  Carroll L. Wainwright and
                  Cary Bassin and
                  Cary Hudson and
                  Casey Chu and
                  Chad Nelson and
                  Chak Li and
                  Chan Jun Shern and
                  Channing Conger and
                  Charlotte Barette and
                  Chelsea Voss and
                  Chen Ding and
                  Cheng Lu and
                  Chong Zhang and
                  Chris Beaumont and
                  Chris Hallacy and
                  Chris Koch and
                  Christian Gibson and
                  Christina Kim and
                  Christine Choi and
                  Christine McLeavey and
                  Christopher Hesse and
                  Claudia Fischer and
                  Clemens Winter and
                  Coley Czarnecki and
                  Colin Jarvis and
                  Colin Wei and
                  Constantin Koumouzelis and
                  Dane Sherburn},
  title        = {GPT-4o System Card},
  journal      = {CoRR},
  volume       = {abs/2410.21276},
  year         = {2024}
}

@inproceedings{LuoFQWLY24,
  author       = {Weilin Luo and
                  Weiyuan Fang and
                  Junming Qiu and
                  Hai Wan and
                  Yanan Liu and
                  Rongzhen Ye},
  title        = {{ITG:} Trace Generation via Iterative Interaction between {LLM} Query
                  and Trace Checking},
  booktitle    = {NIER@ICSE},
  pages        = {11--15},
  year         = {2024}
}

@book{Jersild27,
  title         = {Mental set and shift},
  author        = {Jersild, Arthur Thomas},
  number        = {89},
  year          = {1927}
}

@inproceedings{Liu0L0T25,
  author       = {Yexiang Liu and
                  Jie Cao and
                  Zekun Li and
                  Ran He and
                  Tieniu Tan},
  title        = {Breaking Mental Set to Improve Reasoning through Diverse Multi-Agent
                  Debate},
  booktitle    = {{ICLR}},
  year         = {2025}
}

@inproceedings{Wei0SBIXCLZ22,
  author       = {Jason Wei and
                  Xuezhi Wang and
                  Dale Schuurmans and
                  Maarten Bosma and
                  Brian Ichter and
                  Fei Xia and
                  Ed H. Chi and
                  Quoc V. Le and
                  Denny Zhou},
  title        = {Chain-of-Thought Prompting Elicits Reasoning in Large Language Models},
  booktitle    = {NeurIPS},
  year         = {2022}
}

@inproceedings{ZhengMCCCLZ24,
  author       = {Huaixiu Steven Zheng and
                  Swaroop Mishra and
                  Xinyun Chen and
                  Heng{-}Tze Cheng and
                  Ed H. Chi and
                  Quoc V. Le and
                  Denny Zhou},
  title        = {Take a Step Back: Evoking Reasoning via Abstraction in Large Language
                  Models},
  booktitle    = {{ICLR}},
  year         = {2024}
}

\end{document}